\documentclass[journal,twoside]{IEEEtran}

\usepackage{enumerate}
\usepackage{mathtools}
\usepackage{microtype}
\usepackage{longtable}

\usepackage{pdflscape}

\usepackage{tabularx}
\usepackage{multirow}
\usepackage{textcomp}
\usepackage{csquotes}
\usepackage{booktabs}
\usepackage{longtable}

\usepackage{array}
\usepackage[T1]{fontenc}

\usepackage{xr-hyper}
\usepackage{hyperref}
\usepackage{amsmath}
\usepackage{amssymb}
\usepackage{lscape}
\usepackage{lineno}
\usepackage{array}
\usepackage{float}
\usepackage{todonotes}
\usepackage{tabu}
\usepackage{siunitx}
\usepackage{longtable}
\usepackage{pdflscape}
\usepackage{setspace}
\usepackage{textcomp}
\usepackage{rotating}
\usepackage{xcolor}
\usepackage{acronym}
\usepackage{multicol}
\usepackage{xtab}
\usepackage{multirow}
\usepackage{lipsum} 

\usepackage{xspace}
\usepackage{acronym} 
\usepackage{tablefootnote}
\usepackage{multicol}
\usepackage{float}
\usepackage{threeparttable}

\usepackage{amssymb}
\usepackage{pifont}
\newcommand{\cmark}{\ding{51}}%
\newcommand{\xmark}{\text{\ding{55}}}

\DeclareUnicodeCharacter{2212}{-}

\usepackage{comment}
\def\BibTeX{{\rm B\kern-.05em{\sc i\kern-.025em b}\kern-.08em
    T\kern-.1667em\lower.7ex\hbox{E}\kern-.125emX}}

\ifCLASSINFOpdf
\else
 
\fi

\usepackage[a4paper, total={180mm,257mm}]{geometry}

\begin{document}

%


\title{AI in Thyroid Cancer Diagnosis: Techniques, Trends, and Future Directions}


\author{\IEEEauthorblockN{Yassine Habchi\IEEEauthorrefmark{1},
Yassine Himeur\IEEEauthorrefmark{2}, 
Hamza Kheddar\IEEEauthorrefmark{3}, 
Abdelkrim Boukabou\IEEEauthorrefmark{4,5},
Shadi Atalla\IEEEauthorrefmark{3},
Ammar Chouchane\IEEEauthorrefmark{1},
Abdelmalik Ouamane\IEEEauthorrefmark{6} and
Wathiq Mansoor\IEEEauthorrefmark{3} 
}\\
\IEEEauthorblockA{\IEEEauthorrefmark{1}
Institute of Technology, University Center Salhi Ahmed, BP 58 Naama, 45000 Naama, Algeria; (habchi@cuniv-naama.dz)}\\
\IEEEauthorblockA{\IEEEauthorrefmark{2}College of Engineering and Information Technology, University of Dubai, Dubai, UAE (yhimeur@ud.ac.ae)}\\
\IEEEauthorblockA{\IEEEauthorrefmark{3}LSEA Laboratory, Electrical Engineering Department, University of Medea, 26000, Algeria}\\
\IEEEauthorblockA{\IEEEauthorrefmark{4}Department of Electronics, University of Jijel, BP 98 Ouled Aissa, 18000 Jijel, Algeria}\\
\IEEEauthorblockA{\IEEEauthorrefmark{5}Department of Electrical Engineering, University of Yahia Fares Medea, Algeria}\\
\IEEEauthorblockA{\IEEEauthorrefmark{6}Laboratory of LI3C,  Mohamed Khider University, Biskra,  Algeria}\\
}


\maketitle

\begin{abstract}
There has been a growing interest in creating intelligent diagnostic systems to assist medical professionals in analyzing and processing big data for the treatment of incurable diseases. One of the key challenges in this field is detecting thyroid cancer, where advancements have been made using machine learning (ML) and big data analytics to evaluate thyroid cancer prognosis and determine a patient's risk of malignancy. This review paper summarizes a large collection of articles related to artificial intelligence (AI)-based techniques used in the diagnosis of thyroid cancer. Accordingly, a new classification was introduced to classify these techniques based on the AI algorithms used, the purpose of the framework, and the computing platforms used. Additionally, this study compares existing thyroid cancer datasets based on their features. The focus of this study is on how AI-based tools can support the diagnosis and treatment of thyroid cancer, through supervised, unsupervised, or hybrid techniques. It also highlights the progress made and the unresolved challenges in this field. Finally, the future trends and areas of focus in this field are discussed.
\end{abstract}

\begin{IEEEkeywords}
Thyroid carcinoma detection, Machine learning, Deep learning, Convolutional neural networks.
\end{IEEEkeywords}

\IEEEpeerreviewmaketitle

\section{Introduction}  \label{intro}
\subsection{Background}
The adoption of \ac{AI} in healthcare has become a pivotal development, profoundly reshaping the landscape of medical diagnosis, treatment, and patient care. AI's exceptional capabilities, including pattern recognition, predictive analytics, and decision-making skills, enable the development of systems that can analyze complex medical data at a scale and precision beyond human capacity \cite{himeur2023face}. This, in turn, augments early disease detection, facilitates accurate diagnoses, and aids personalized treatment planning. Moreover, AI-driven predictive models can forecast disease outbreaks, enhance the efficiency of hospital operations, and significantly improve patient outcomes \cite{himeur2022deep}. Additionally, AI has the potential to democratize healthcare by bridging the gap between rural and urban health services and making high-quality care more accessible. Hence, the importance of AI in healthcare is profound and will continue to grow as technology advances, leading to more sophisticated applications and better health outcomes for patients worldwide \cite{sohail2023decoding,himeur2023ai}.

\begin{table*}[]
{\small \textbf{Abbreviations}}
\begin{multicols}{3}
\footnotesize
\begin{acronym}[AWGN]  
\acro{AI}{Artificial intelligence}
\acro{ATC}{Anaplastic thyroid carcinoma}
\acro{ACR TI-RADS}{American college of radiology has formulated a thyroid imaging, reporting, and data system}
\acro{ANN}{Artificial neural network} 
\acro{APP}{Application}
\acro{AC}{Active contour}
\acro{ACC}{Accuracy}
\acro{AI-TS}{AI in thyroid surgery}
\acro{Bi-LSTM}{Bi-Long Short Term Memory}
\acro{BN}{Bayesian networks}
\acro{B}{Bagging}
\acro{BA}{Bootstrap aggregation}
\acro{BPSO}{Binary particle swarm optimization}
\acro{CAD}{Computer-aided diagnosis}
\acro{CNN}{Convolutional neural network}
\acro{C}{Clustering}
\acro{CFS}{Correlation-based feature selection}
\acro{CORR}{Correlation} 
\acro{CT}{Computed tomography} 
\acro{DL}{Deep learning}
\acro{DNN}{Deep neural network}  
\acro{DAE}{Denoising autoencoder}
\acro{DT}{Decision trees}
\acro{DR}{Dimensionality reduction}
\acro{DTCW}{Double-tree complex wavelet transform}
\acro{DD}{Detected diseas}
\acro{DDTI}{Digital database thyroind image}
\acro{DA}{Dataset access}
\acro{DCG}{Discounted cumulative gain}
\acro{ELM}{Extreme learning machine} 
\acro{EB}{Entropy-based}
\acro{EM}{Ensemble methods}
\acro{ER}{Error Rate}
\acro{EFC-AI}{Edge, fog, and cloud networks based on AI}
\acro{FTC}{Follicular thyroid carcinoma}
\acro{FNAB}{Fine-needle aspiration biopsy}
\acro{FKNN}{Fuzzy K-nearest neighborhood}
\acro{FB}{Feature bagging}
\acro{FS}{Feature selection methods}
\acro{F}{Female}
\acro{FE}{Feature extraction methods}
\acro{F1}{F1 Score}
\acro{FPR}{Fallout or false positive rate}
\acro{FN}{False negative}
\acro{FP}{False positive}
\acro{FL}{Federated learning}
\acro{GAN}{Generative adversarial network}
\acro{GWAS}{Genome-wide association analysis} 
\acro{GEO}{Gene expression omnibus}
\acro{GLCM}{Gray-level co-occurrence matrix}
\acro{JSI}{Jaccard similarity index} 
\acro{HTC}{Hürthle thyroid carcinoma}
\acro{HOG}{Histogram of oriented gradient}
\acro{IG}{Information gain}
\acro{IT}{Image types}
\acro{IF}{Image format}
\acro{ICA}{Independent component analysis}
\acro{IoMIT}{Internet of medical imaging thing}
\acro{KNN}{k-nearest neighbors}
\acro{KM}{K-means}
\acro{LSTM}{long-short-term-memory}
\acro{LTC}{Lymphoma thyroid carcinoma}
\acro{LR}{Logistic regression}
\acro{LBP}{Local binary patterns}
\acro{LMSE}{Laplacian mean square error}
\acro{MLP}{Multilayer perceptron}
\acro{ML}{Machine learning}
\acro{MTC}{Medullary thyroid carcinoma}
\acro{M}{Male}
\acro{MAE}{Mean absolute error}
\acro{MSE}{Mean squared error}
\acro{MRR}{Mean reciprocal rank}
\acro{MRI}{Magnetic resonance imaging}
\acro{NP}{ Number of patients}
\acro{NM}{Number of males}
\acro{NF}{Number of females}
\acro{NN}{Number of nodules}
\acro{NBN}{Number of benign nodules}
\acro{NMN}{Number of malignant Nodule}
\acro{NCDR}{National cancer data repository}
\acro{NCI}{National cancer institute}
\acro{NPV}{Negative predictive value}
\acro{NCC}{Normalized cross-correlation} 
\acro{NVF}{Noise visibility function}
\acro{NAE}{Normalized absolute error}
\acro{OB}{Objective}
\acro{PTC}{Papillary carcinoma}
\acro{PM}{Probabilistic models}
\acro{PCA}{Principal component analysis}
\acro{PD}{Private data}
\acro{PLCO}{Prostate, lung, colorectal, and ovarian} 
\acro{P}{Precision}
\acro{PSNR}{Peak signal to noise ratio} 
\acro{PS}{Panoptic segmentation}
\acro{RNN}{Recurrent neural network}
\acro{RBM}{Restricted boltzmann machine}
\acro{RBF}{Radial basis function} 
\acro{RF}{Random forest}
\acro{R}{Relief}
\acro{RMSE}{Root mean square error} 
\acro{RL}{Reinforcement learning}
\acro{RS}{Recommender systems}
\acro{SVM}{Support vector machine}
\acro{SL}{Supervised learning} 
\acro{STC}{Sarcoma thyroid carcinoma}
\acro{SV}{Subjects for validation}
\acro{SEER}{Surveillance, epidemiology, and end results}
\acro{SPE}{Specificity}
\acro{SEN}{Sensitivity}
\acro{SD}{Standard deviation}
\acro{SSIM}{Structural similarity index}
\acro{SC}{Structural content}
\acro{SIFT}{Sscale invariant feature transform} 
\acro{TN}{Thyroid nodules}
\acro{TG}{Thyroid gland} 
\acro{TD}{Thyroid disease} 
\acro{TCL}{Traditional classification}
\acro{TMC}{Thyroid microcarcinoma} 
\acro{TDDS}{Thyroid disease data set} 
\acro{TCD}{Thyroid cancer datase}
\acro{TCGA}{Cancer genome atlas}
\acro{TD}{Texture description}
\acro{TI}{Thyroid inputs}
\acro{TT}{Thyroid Targets} 
\acro{TS}{Thyroid segmentation} 
\acro{TL}{Transfer learning}
\acro{TN}{True negative}
\acro{TP}{True positive}
\acro{TIRADS}{Thyroid imaging reporting and data system}
\acro{US}{Ultrasound}
\acro{USL}{Unsupervised learning}
\acro{WEKA}{Waikato environment for knowledge analysis}
\acro{WBCD}{Wisconsin breast cancer dataset}
\acro{WP-SNR}{Weight peak signal to noise ratio}
\acro{WSNR}{Weighted signal-to-noise ratio}
\acro{WSI}{Whole slide images}
\acro{VOE}{Volumetric overlap error} 
\acro{VIF}{Visual information fidelity} 
\acro{VSNR}{Visual signal to noise ratio}
\acro{XAI}{Explainable artificial intelligence}
\acro{3D-TCD}{3D thyroid cancer detection} 
\end{acronym}
\end{multicols}
\end{table*}

\begin{figure*}[t!]
\includegraphics[width=2\columnwidth]{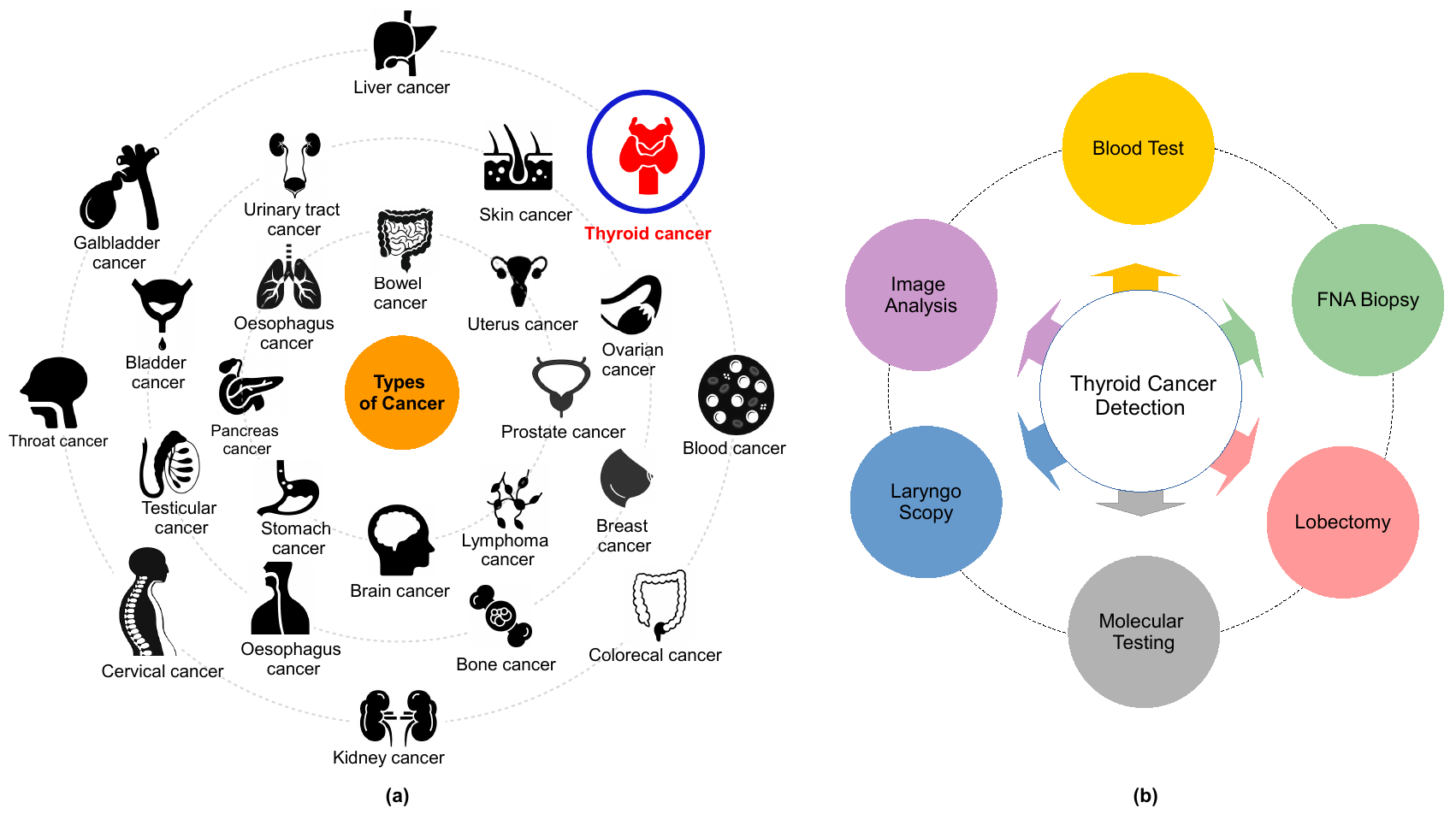}
\caption{(a) The different kinds of cancer and (b) Thyroid cancer detection methods.}
\label{fig2}
\end{figure*}

Cancer, a leading cause of death, affects various parts of the body as depicted in Fig. \ref{fig2} (a). Among various types, thyroid carcinoma stands out as one of the most commonly occurring endocrine cancers globally \cite{deng2020global, hammouda2006registre}. Concerns are mounting over the escalating incidence of thyroid cancer and associated mortality. Research indicates that thyroid cancer incidence is higher in women aged 15-49 years (ranked fifth globally) compared to men aged 50-69 years \cite{abid2008cancer, NIPH2021, siegel2019cancer}.

According to existing global epidemiological data, the rapid growth of abnormal thyroid nodules (TN) is driven by an accelerated increase in genetic cell activity. This condition can be categorized into four primary subtypes: \ac{PTC} \cite{hitu2020microrna}, follicular carcinoma (FTC) \cite{castellana2020can}, anaplastic carcinoma (ATC) \cite{ferrari2020novel}, and medullary carcinoma (MTC) \cite{giovanella2020eanm}. Influential factors such as high radiation exposure, Hashimoto's thyroiditis, psychological and genetic predispositions, as well as advancements in detection technology, can contribute to the onset of these cancer types. These conditions might subsequently lead to chronic health issues, including diabetes, irregular heart rhythms, and blood pressure fluctuations \cite{carling2014thyroid, yang2020comparison, kobayashi2018calcifications}.
Although the quantity of cancer cells is a significant indicator of thyroid carcinoma, obtaining results is often time-consuming due to the requirement to observe cell appearance. Thus, the detection and quantification of cell nuclei are considered crucial biomarkers for assessing cancer cell proliferation.

The utilization of computer-aided diagnosis (CAD) systems for analyzing thyroid cancer images has seen a significant increase in popularity in recent years. These systems, renowned for enhancing diagnostic accuracy and reducing interpretation time, have become an invaluable tools in the field. Among these technologies, radionics, when used in conjunction with ultrasonography imaging, has become widely accepted as a cost-effective, safe, simple, and practical diagnostic method in clinical practice.
Endocrinologists frequently conduct US scans in the 7-15 MHz range to identify thyroid cancer and evaluate its anatomical characteristics. The American College of Radiology has formulated a Thyroid Imaging, Reporting, and Data System (ACR TI-RADS) that classifies thyroid nodules into six categories based on attributes like composition, echogenicity, shape, size, margins, and echogenic foci. These classifications range from normal (TIRADS-1) to malignant (TIRADS-6) \cite{tessler2017acr, tessler2018thyroid, ABCD1}. Several open-source applications are available for assessing these thyroid cancer features \cite{AB1, ABC1}.
However, the identification and differentiation of nodules continue to present a challenge, largely reliant on the personal experience and cognitive abilities of radiologists. This is due to the subjective nature of human eye-based image recognition, the poor quality of captured images, and the similarities among US images of benign thyroid nodules, malignant thyroid nodules, and lymph nodes.

Moreover, ultrasonography imaging is often a time-intensive and stressful procedure, which can result in inaccurate diagnoses. Misclassifications among normal, benign, malignant, and indeterminate cases are common \cite{schlumberger2015lenvatinib, wettasinghe2019diagnostic, nayak2020impact, singh2016diagnostic, kumar2020automated, ma2017ultrasound}. For a more precise diagnosis, a fine-needle aspiration biopsy (FNAB) is typically conducted. However, FNAB can be an uncomfortable experience for patients, and a specialist's lack of experience can potentially convert benign nodules into malignant ones, not to mention the additional financial burden \cite{song2018fine, hahn2020comparison} (refer to Fig.\ref{fig2} (b)).
The primary challenge in distinguishing between benign and malignant nodules resides in the selection of their characteristics. Numerous studies have explored the characterization of conventional US imaging for various types of cancers, including retina \cite{ullah2019ensemble, saba2018fundus}, breast cancer \cite{mughal2018removal, mughal2018novel}, blood cancer \cite{abbas2019automated, abbas2019plasmodium}, and thyroid cancer \cite{wang2020comparison, qin2019diagnosis}. However, these methods still fall short when it comes to the accurate classification of thyroid nodules.

The incorporation of AI technology plays a pivotal role in reducing subjectivity and enhancing the accuracy of pathological diagnoses for various intractable diseases, including those affecting the thyroid gland \cite{wu2016classifier, zhang2019machine}. This enhancement is achieved through improved interpretation of ultrasonography images and faster processing times. Machine learning (ML) and deep learning (DL) have surfaced as potential solutions for automating the classification of thyroid nodules in applications such as US, fine-needle aspiration (FNA), and thyroid surgery \cite{sollini2018texture, yang2019creating}. This potential has been underscored in numerous studies, such as \cite{abbad2021effective, taylor2019high, chandio2020decision, zhang2019machine, lee2021deep, buda2019management}.
Furthermore, there are ongoing studies examining the use of this innovative technology for cancer detection, where its effectiveness hinges on the volume of data and the precision of the classification process.

The motivation to write a review on "Artificial Intelligence for Detecting Thyroid Carcinoma" stems from the increasing prevalence of thyroid cancer, a significant endocrine malignancy where early and accurate detection is pivotal for patient outcomes. As technological advancements in AI and machine learning burgeon, their integration into medical diagnostics—spanning imaging, pathology, and genomics—offers potential improvements in detection accuracy and efficiency. Traditional thyroid carcinoma diagnostic methods, like fine-needle aspiration biopsies, sometimes present inconclusive results; AI promises less invasive alternatives with possibly superior precision. Such a review would amalgamate insights from the intersection of computer science, radiology, pathology, and endocrinology, propelling multidisciplinary collaboration. It would also spotlight AI's clinical implications, guiding clinicians in leveraging its capabilities for patient care, while delineating future research directions. Furthermore, this review would underscore the economic and healthcare benefits, from cost savings to reduced waiting times. At the same time, it is imperative to address AI's inherent challenges, including data privacy and ethical considerations, ensuring its balanced integration into healthcare. In essence, the review would offer a comprehensive panorama of AI's current and potential role in thyroid carcinoma detection, benefitting both researchers and medical practitioners.

\subsection{Our contribution}
This review provides a comprehensive examination of the application of Artificial Intelligence (AI) methods in detecting thyroid cancer. The objective of AI-based analysis in the medical field is increasingly shifting towards enhancing diagnostic accuracy, and this review aims to illustrate this trend, particularly in thyroid cancer detection.
We first provide an overview of the existing frameworks and delve into the specifics of various AI techniques. These include supervised learning methods, like DL, artificial neural networks, traditional classification, and probabilistic models, as well as unsupervised learning methods, such as clustering and dimensionality reduction. We also explore ensemble methods, including bagging and boosting.
Recognizing the importance of quality datasets in AI applications, we scrutinize several thyroid cancer datasets, addressing their features, as well as feature selection and extraction methods used in various studies. We then outline the standard assessment criteria used to evaluate the performance of AI-based thyroid cancer detection methods. These range from classification and regression metrics to statistical metrics, computer vision metrics, and ranking metrics.
Finally, we discuss future research directions, emphasizing areas that require more attention to overcome existing barriers and improve the use and deployment of thyroid cancer detection solutions. In conclusion, we underscore the potential of AI in advancing thyroid cancer detection while also noting the need for continuous critical evaluation to ensure its responsible and effective use.

All in all, the principal contributions of our paper are as follows
\begin{itemize}
\item Overview of existing frameworks and specifics of various AI techniques, including supervised learning (DL, artificial neural networks, traditional classification, and probabilistic models) and unsupervised learning (clustering and dimensionality reduction) methods, as well as ensemble methods (bagging and boosting).
\item Scrutiny of several thyroid cancer datasets, addressing their features, feature selection, and extraction methods used in various studies.
\item Outline of standard assessment criteria used to evaluate the performance of AI-based thyroid cancer detection methods, encompassing classification and regression metrics, statistical metrics, computer vision metrics, and ranking metrics.
\item Critical analysis and discussion highlighting limitations, hurdles, current trends, and open challenges in the field.
\item Discussion of future research directions, emphasizing areas requiring more attention to overcome existing barriers and improve thyroid cancer detection solutions.
\item Emphasis on the potential of AI in advancing thyroid cancer detection while advocating continuous critical evaluation for responsible and effective use.

\end{itemize}

Additionally, the principal contributions of the proposed review compared to other existing surveys are summarized in Table \ref{table:1}.

\begin{table*}[t!]
\caption{The major contributions of the proposed contributions on thyroid cancer classification in comparison with other related works.}
\label{table:1}
\begin{center}

\begin{tabular}{ccccccc}
\hline
Ref & Year & PPY & TCDS & AIA 

& 
\begin{tabular}{c}
Open challenges \\ \hline
\multicolumn{1}{c}{%
\begin{tabular}{ccc}
TCDA & RDLA & PP%
\end{tabular}%
}%
\end{tabular}
& 

\begin{tabular}{c}
Future directions \\ \hline
\multicolumn{1}{c}{%
\begin{tabular}{cccccc}
XAI & EFC-AI & RL & PS & IoMIT & RS%
\end{tabular}%
}%
\end{tabular}
\\ \hline

\cite{liu2021deep} & 2021 & \cmark & \cmark & \xmark & 
\begin{tabular}{ccc}
\xmark & \xmark & \xmark%
\end{tabular}
& 
\begin{tabular}{cccccc}
\xmark & \xmark & \xmark & \xmark & \xmark & \xmark%
\end{tabular}
\\ 
\cite{iesato2021role} & 2021 & \cmark & \cmark & \xmark & 
\begin{tabular}{ccc}
\xmark & \xmark & \xmark%
\end{tabular}
& 
\begin{tabular}{cccccc}
\xmark & \xmark & \xmark & \xmark & \xmark & \xmark%
\end{tabular}
\\ 
\cite{sharifi2021deep} & 2021 & \cmark & \cmark & \xmark & 
\begin{tabular}{ccc}
\xmark & \xmark & \xmark%
\end{tabular}
& 
\begin{tabular}{cccccc}
\xmark & \xmark & \xmark & \xmark & \xmark & \xmark%
\end{tabular}
\\ 
\cite{lin2021deep} & 2021 & \cmark & \cmark & \xmark & 
\begin{tabular}{ccc}
\xmark & \xmark & \xmark%
\end{tabular}
& 
\begin{tabular}{cccccc}
\xmark & \xmark & \xmark & \xmark & \xmark & \xmark%
\end{tabular}
\\ 
\cite{ha2021applications} & 2021 & \cmark & \cmark & \xmark & 
\begin{tabular}{ccc}
\xmark & \xmark & \xmark%
\end{tabular}
& 
\begin{tabular}{cccccc}
\xmark & \xmark & \xmark & \xmark & \xmark & \xmark%
\end{tabular}
\\ 
\cite{wu2022deep} & 2022 & \cmark & \cmark & \xmark & 
\begin{tabular}{ccc}
\xmark & \xmark & \xmark%
\end{tabular}
& 
\begin{tabular}{cccccc}
\xmark & \xmark & \xmark & \xmark & \xmark & \xmark%
\end{tabular}
\\ 
\cite{pavithra2022deep} & 2022 & \cmark & \cmark & \xmark & 
\begin{tabular}{ccc}
\xmark & \xmark & \xmark%
\end{tabular}
& 
\begin{tabular}{cccccc}
\xmark & \xmark & \xmark & \xmark & \xmark & \xmark%
\end{tabular}
\\ 
\cite{paul2022artificial} & 2022 & \cmark & \cmark & \xmark & 
\begin{tabular}{ccc}
\xmark & \xmark & \xmark%
\end{tabular}
& 
\begin{tabular}{cccccc}
\xmark & \xmark & \xmark & \xmark & \xmark & \xmark%
\end{tabular}
\\ 
\cite{ilyas2022deep} & 2022 & \cmark & \cmark & \xmark & 
\begin{tabular}{ccc}
\xmark & \xmark & \xmark%
\end{tabular}
& 
\begin{tabular}{cccccc}
\xmark & \xmark & \xmark & \xmark & \xmark & \xmark%
\end{tabular}
\\ 
Our & - & \cmark & \cmark & \cmark & 
\begin{tabular}{ccc}
\cmark & \cmark & \cmark%
\end{tabular}
& 
\begin{tabular}{cccccc}
\cmark & \cmark & \cmark & \cmark & \cmark & \cmark%
\end{tabular}
\\ \hline
\end{tabular}%

\end{center}
\end{table*}

\subsection{Roadmap}
The rest of this paper is organized as follows. Section II follows, providing an overview of existing frameworks utilized in this field, and discussing their respective advantages and limitations. Section III presents various thyroid cancer datasets used in AI-based analyses, explaining their relevance and uniqueness. In Section IV, the paper delves into the vital aspect of 'Features', discussing feature extraction and selection methods in AI models used for thyroid cancer detection. Section V outlines the standard assessment criteria used to evaluate the performance of these models. An actual instance of AI-based thyroid cancer detection is presented in Section VI to provide a real-world context to the theoretical aspects discussed earlier. The paper then proceeds to critical analysis and discussion in Section VII, where challenges, limitations, and areas for improvement in the current approaches are discussed. In Section VIII, potential future research directions are proposed, highlighting areas where further exploration and innovation can lead to advancements in AI-based thyroid cancer detection. The paper concludes with Section IX, summarizing the main findings and discussions, thereby providing a comprehensive conclusion to the discussions presented in the earlier sections.


\section{Overview of existing frameworks}
\begin{figure*}[t!]

\includegraphics[width=2\columnwidth]{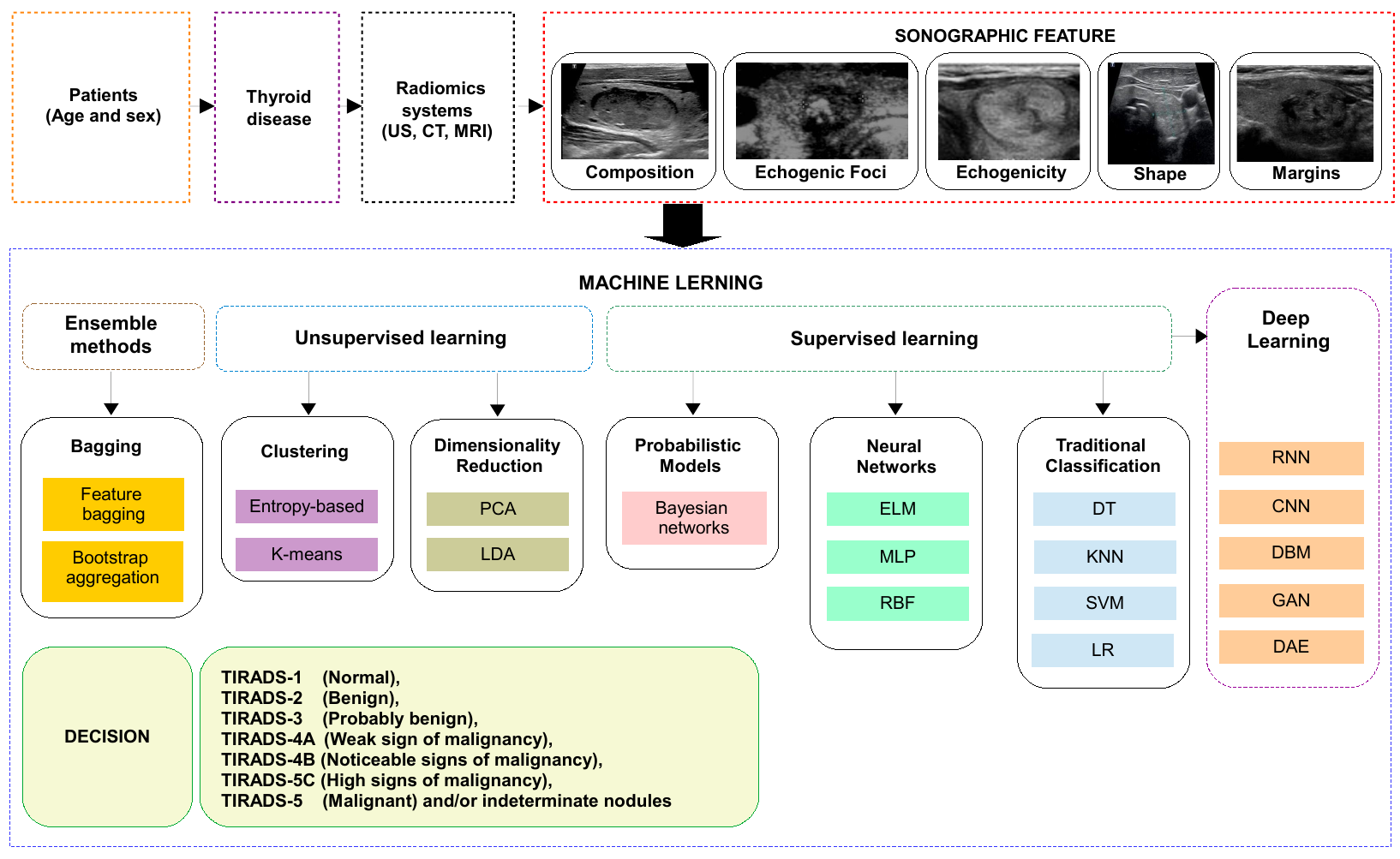}
\caption{Taxonomy of the thyroid cancer detection schemes based in AI.}
\label{fig3}

\end{figure*}

This section showcases the various AI-based methods utilized for diagnosing thyroid gland (TG) cancers. In the illustration, Fig. \ref{fig3} presents a proposed categorization of the thyroid cancer diagnosis techniques relying on AI.

\subsection{Objective of AI-based analysis (O)}
This article focuses on the application of AI in thyroid cancer detection. In order to better understand the purpose behind each framework, it is crucial to identify the objective of each approach.
\vskip2mm
\noindent \textbf{O1. Classification}: Thyroid carcinoma classification refers to the categorization of thyroid cancers based on their histopathological features, clinical behavior, and prognosis. There are several types of thyroid carcinomas, each of which has distinct characteristics. The primary categories include: (i) Papillary Thyroid Carcinoma (PTC): The most common type, accounting for about 80\% of all thyroid cancers. PTC tends to grow very slowly, but it often spreads to lymph nodes in the neck. Despite this, it is usually curable with treatment; (ii) Follicular Thyroid Carcinoma (FTC): The second most common type, FTC can invade blood vessels and spread to distant parts of the body, but it is less likely to spread to lymph nodes; (iii) Medullary Thyroid Carcinoma (MTC): This type of thyroid cancer starts in the thyroid's parafollicular cells, also called C cells, which produce the hormone calcitonin. Elevated levels of calcitonin in the blood can indicate MTC; and (iv) Anaplastic Thyroid Carcinoma (ATC): A very aggressive and rare form of thyroid cancer, ATC often spreads quickly to other parts of the neck and body. It is difficult to treat.

The classification of thyroid carcinomas is crucial in determining the most effective course of treatment for each patient. Various factors such as tumor size, location, and the patient's age and overall health are also taken into consideration when forming a treatment plan. Advances in AI and machine learning are helping to automate and improve the accuracy of thyroid carcinoma classification, with many models trained to classify tumors based on medical images or genetic data.
As reported by Liu et al. \cite{liu2018setsvm}, incorporates \ac{SVM} for cancer detection. Similarly, Zhang et al. \cite{zhang2019integrating,zhang2019diagnosis} propose \ac{DNN} based strategies for segregating and categorizing benign and malignant thyroid nodules in ultrasound imagery. Furthermore, the Bi-Long Short Term Memory (Bi-LSTM) model, as presented by Chen et al. \cite{chen2018thyroid}, demonstrates notable accuracy in classifying thyroid nodules. These classification systems constitute structured hierarchies instrumental in organizing knowledge and workflows in the specific domain of thyroid cancer.

\vskip2mm
\noindent \textbf{O2. Segmentation}: segmentation of thyroid carcinoma refers to the process of identifying and delineating the region of an image that corresponds to a thyroid tumor.
The goal of segmentation is to separate the areas of interest, in this case, the thyroid tumor, from the surrounding tissues in the medical images. This can be done manually by an expert radiologist, or it can be automated using machine learning algorithms \cite{ma2023amseg, yadav2023assessment}.
Segmentation is a crucial step in medical image analysis because it helps to accurately determine the location, size, and shape of the tumor, which are vital parameters for diagnosis, treatment planning, and prognosis prediction.
A variety of methods can be used to perform image segmentation, including thresholding, edge detection, region-growing methods, and more complex machine learning and DL techniques.

In the case of thyroid carcinoma, the segmentation can be challenging due to the high variability in the appearance and shape of the tumors, their proximity to other structures in the neck, and the presence of noise or artifacts in the images. Therefore, robust and reliable segmentation algorithms are needed to ensure accurate and consistent results.
AI methods, including \ac{CNN} and U-Net architecture, are being increasingly used for thyroid carcinoma segmentation because of their ability to learn and generalize from large amounts of data, thus improving the accuracy and reliability of the segmentation process.


\vskip2mm
\noindent \textbf{O3. Prediction}: The prediction of thyroid carcinoma involves the use of various diagnostic tools, tests, and techniques - often employing machine learning models - to anticipate the probability of a patient developing thyroid cancer. This predictive analysis can be based on several factors, including but not limited to (i) Genetic predisposition: Individuals with a family history of thyroid cancer are at a higher risk; (ii) Gender and age: Thyroid cancer is more common in women and people aged between 25 and 65; (iii) Radiation exposure: Exposure to high levels of radiation, especially during childhood, increases the risk of developing thyroid cancer; (iv) Diet and lifestyle: Lack of iodine in the diet and certain lifestyle factors may contribute to an increased risk.
In a medical context, prediction does not necessarily mean a certain future outcome, but rather it points to an increased risk or likelihood based on current data and predictive models. For thyroid carcinoma, predictive tools and tests are typically used in conjunction with each other to achieve more accurate results. For instance, machine learning algorithms can be trained on historical medical data to predict the likelihood of a nodule being benign or malignant, aiding in early detection and more effective treatment planning.
Various studies have been proposed to predict thyroid cancer. For instance, in \cite{jajroudi2014prediction}, the authors employed the use of Artificial Neural Network (ANN) and Logistic Regression to make predictions. Another study \cite{sajeev2020thyroid} details the creation of a predictive machine using Convolution Neural Networking (CNN) to analyze 10068 microscopic thyroid cancer images from South Asian populations.

\subsection{Supervised learning (SL)}
Supervised learning is a method of machine learning where an algorithm is trained to classify or predict the condition based on labeled data, which in this case is medical data related to thyroid cancer. The aim of supervised learning is to differentiate between the different forms of thyroid cancer through the use of annotated data and examples.
For example, this data can include ultrasound images, radiomic features, genetic markers, patient demographics, or any other information that may be relevant to the diagnosis or prognosis of thyroid cancer. The labeled data would indicate whether each instance corresponds to a case of thyroid cancer or not, or it may provide more detailed labels such as the stage of the cancer or the type of thyroid carcinoma.

In a classification setting, the supervised learning algorithm could be trained to distinguish between benign and malignant thyroid nodules based on certain characteristics extracted from medical imaging data. The labels in the training data would specify whether each nodule is benign or malignant. After training, the algorithm can then be used to classify new, unlabeled nodules.
Similarly, a regression-based supervised learning algorithm might be trained to predict the progression or the prognosis of thyroid cancer based on various patient-specific features. The labels here would correspond to a continuous outcome variable, such as the survival time of the patient or a measure of disease progression.
It is important to note that the performance of these methods heavily relies on the quality and quantity of the available data. The more accurate and comprehensive the data, the better the algorithm will perform in predicting or classifying new instances. Additionally, supervised learning models in healthcare, including thyroid carcinoma detection, need to be validated on separate test datasets and in real-world clinical settings to ensure their robustness and reliability \cite{liu2023self, hou2023boosting}.

\subsubsection{Deep learning (DL)}
\begin{figure*}[t!]
\includegraphics[width=2\columnwidth]{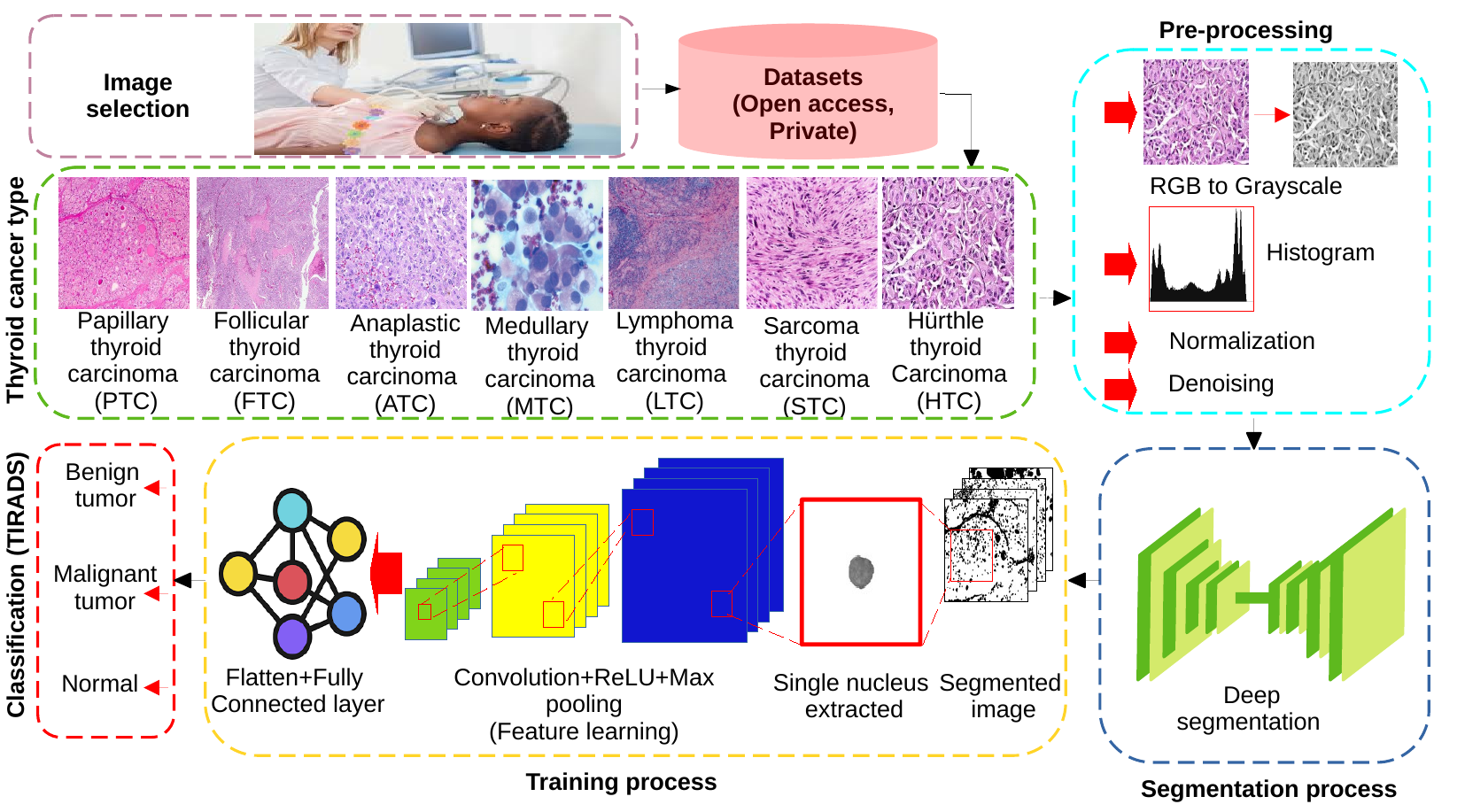}
\caption{General DL system for thyroid cancer detection and classification.}
\label{fig4}
\end{figure*}

DL is a subset of machine learning and artificial intelligence that's based on artificial neural networks with representation learning. It can automatically learn, generate, and improve representations of data by employing large neural networks with many layers—hence the term "deep" learning.
In thyroid cancer, DL has been deployed to perform different tasks, including (i) Image Classification -- DL algorithms like Convolutional Neural Networks (CNNs) can be trained to classify thyroid ultrasound images. For instance, they can differentiate between benign and malignant nodules based on their shape, texture, and other characteristics  \cite{canton2021automatic, peng2021deep, guan2019deep}. This approach can significantly reduce the time and effort required for manual interpretation, thus aiding in the early detection and treatment of thyroid cancer; (ii) Pathological Analysis -- DL can also be utilized to analyze histopathological or cytopathological slide images, helping in the detection and classification of cancerous cells; (iii) Genomic Data Analysis -- With the advent of genomic medicine, DL models can be employed to analyze genetic variations that may predispose individuals to thyroid cancer; (iv)  Radiomics -- DL models can be used to extract high-dimensional data from radiographic images, allowing for more precise and personalized treatment planning; and (v) Predictive Analysis -- Using electronic health records and other patient data, DL models can be used to predict the likelihood of a patient developing thyroid carcinoma, allowing for preventive measures to be taken if necessary. 
Fig. \ref{fig4} illustrates the different classifications of thyroid cancer using DNN.

\vskip2mm

\noindent \textbf{D1. Denoising autoencoder (DAE):}
Denoising autoencoders (DAEs) can be beneficial for thyroid carcinoma classification by effectively learning representations from ultrasound or histopathological images. A DAE is a specific type of artificial neural network trained to reconstruct input data, often used for the purposes of dimensionality reduction or feature learning. The process for utilizing DAEs for thyroid carcinoma classification generally follows these steps (i) preprocessing, (ii) noisy input creation, (iii) DAE training, (iv) feature extraction and (v) classification.
In \cite{ferreira2018autoencoders}, the authors implemented six autoencoder algorithms in the training process for papillary thyroid carcinoma (PTC) classification, including fixing weights and fine-tuning the network. The encoding layers and the complete auto-encoder were used to embed the network. Another study \cite{teixeira2017learning} employed denoising autoencoders (DAE) and stacked denoising autoencoders to extract features and identify informative genes in thyroid cancer.

\vskip2mm

\noindent \textbf{D2. CNN:}
CNNs are a class of DL models that have shown extraordinary performance in various image processing and analysis tasks, including the classification of medical images. CNNs are especially adept at processing grid-like data, such as an image, where spatial relationships between the pixels are crucial to understanding the image content.
The past few years have seen considerable effort invested in developing CNN-based methodologies for detecting thyroid cancer, especially for the automated identification and classification of nodules in ultrasound imagery \cite{liu2019automated}. The ConvNet model, a widely adopted framework within the neural network realm, emphasizes the use of convolution operations over matrix multiplications \cite{ha2019deep}. Various CNN architectures such as LeNet \cite{zhang2020detection}, AlexNet \cite{qiao2021deep}, VGG \cite{guan2019deep}, ResNet \cite{cox2020hyperparameter}, GoogLeNet \cite{chi2017thyroid}, Squeeze Net \cite{tekchandani2021severity}, and DenseNet \cite{ke2020development}, are distinguished by their incorporation of key components including convolutional, pooling, and fully connected layers.

In a study conducted by \cite{li2019diagnosis}, the potential of CNN models to prognosticate thyroid cancer was explored using 131,731 ultrasound images taken from 17,627 thyroid cancer patients. Another research effort \cite{xie2019thyroid} employed VGG16, Inception, and Inception-Resnet models to differentiate malignant tissues within a set of 451 thyroid images from the DDTI dataset. To mitigate the challenge of data scarcity, the images were augmented before classification. A comparison of DCNN diagnostic performance with expert radiologists in distinguishing thyroid nodules (TN) within ultrasound images was carried out by \cite{koh2020diagnosis}, involving a test set of 15,375 TN ultrasound images. They utilized CNNE1 and CNNE2 models, derived from DCNN, for differentiating between malignant and benign TNs. The study \cite{liang2020convolutional} proposed a CNN-based DL technique for detecting and classifying TN and breast nodules, with the results contrasted against those from ultrasound imaging.
Table \ref{table:2} presents a summary of recent CNN-based thyroid cancer classification contributions.

\begin{table*}[!t]
\caption{Summary of CNN research conducted in the diagnosis of thyroid cancer.}
\label{table:2}

\scriptsize
\begin{tabular}{
m{1cm}
m{1cm}
m{1cm}
m{1cm}
m{1cm}
m{1cm}
m{1cm}
m{1cm}
m{1cm}
m{1cm}
m{1cm}
m{1cm}
m{1cm}
}
\hline
Ref. & Year & Country & NP & NM & NF &  NN & NBN & NMN & TP & TN & FP & FN \\ \hline


{\small \cite{zhu2021efficient}} & {\small 2021} & {\small China} & {\small 102} & {\small 00} & {\small 102} & {\small 104} & {\small 57} & {\small 47} & {\small 38} & {\small 07} & {\small 07} & {\small 50} \\ 

{\small \cite{zhao2021comparative}} & {\small 2021} & {\small China } & {\small 102} & {\small 25} & {\small 77} & {\small 103} & {\small 73} & {\small 33} & {\small 27} & {\small 12} & {\small 06} & {\small 61} \\ 

{\small \cite{han2021computer}} & {\small 2021} & {\small Koria} & {\small 325} & {\small 61} & {\small 264} & {\small 325} & {\small 257} & {\small 68} & {\small 48} & {\small 52} & {\small 20} & {\small 205} \\

{\small \cite{liu2021thyroid}} & {\small 2021} & {\small China } & {\small 2775} & {\small 726} & {\small 2049} & {\small 2775} & {\small 2472} & {\small 303} & {\small 271} & {\small 363} & {\small 32} & {\small 2109} \\

{\small \cite{liu2021thyroid}} & {\small 2021} & {\small China} & {\small 163} & {\small 48} & {\small 115} & {\small 175} & {\small 67} & {\small 108} & {\small 86} & {\small 09} & {\small 22} & {\small 58} \\

{\small \cite{wei2020value}} & {\small 2020} & {\small China} & {\small 2489} & {\small 614} & {\small 1875} & {\small 2489} & {\small 1021} & {\small 1468} & {\small 1280} & {\small 258} & {\small 188} & {\small 763} \\

{\small \cite{stib2020thyroid}} & {\small 2020} & {\small USA} & {\small 571} & {\small 234} & {\small 337} & {\small 651} & {\small 500} & {\small 151} & {\small 133} & {\small 214} & {\small 18} & {\small 287} \\

{\small \cite{ye2020intelligent}} & {\small 2020} & {\small China} & {\small 166} & {\small 46} & {\small 100} & {\small 209} & {\small 109} & {\small 100} & {\small 87} & {\small 16} & {\small 13} & {\small 93} \\

{\small \cite{koh2020diagnosis}} & {\small 2020} & {\small Korea} & {\small 200} & {\small 49} & {\small 151} & {\small 200} & {\small 102} & {\small 98} & {\small 90} & {\small 41} & {\small 08} & {\small 61} \\

{\small \cite{shin2020application}} & {\small 2020} & {\small Korea} & {\small 340} & {\small 79} & {\small 261} & {\small 348} & {\small 252} & {\small 96} & {\small 31} & {\small 25} & {\small 65} & {\small 227} \\

{\small \cite{park2019diagnosis}} & {\small 2019} & {\small Korea} & {\small 106} & {\small 29} & {\small 77} & {\small 2018} & {\small 132} & {\small 86} & {\small 69} & {\small 23} & {\small 17} & {\small 109} \\

{\small \cite{xia2019computer}} & {\small 2019} & {\small China} & {\small 171} & {\small 32} & {\small 139} & {\small 180} & {\small 85} & {\small 95} & {\small 86} & {\small 50} & {\small 09} & {\small 35} \\

\\ \hline
\end{tabular}%
\par
\begin{tablenotes}
\item Abbreviations: Number of patients (NP); Number of males (NM); Number of females (NF); Number of nodules (NN); Number of benign nodules (NBN); Number of malignant Nodules (NMN).
\end{tablenotes}
\end{table*}

\vskip2mm

\noindent \textbf{D3. Recurrent neural network (RNN):}
Recurrent Neural Networks (RNNs) are a class of artificial neural networks where connections between nodes form a directed graph along a sequence, thus enabling them to use their internal state (memory) to process variable-length sequences of inputs. This unique feature makes RNNs particularly suitable for tasks where temporal dependencies are essential, such as time-series analysis, language translation, and speech recognition.
In the context of thyroid carcinoma classification, RNNs can be utilized to analyze sequential or time-dependent data, such as the development of a patient's clinical signs over time, the evolution of a tumor seen in a series of medical images, or changes in the gene expression related to the progression of thyroid cancer.
For instance, in the study by Chen et al. (2017) \cite{chen2017thyroid}, the authors propose a hierarchical Recurrent Neural Network (RNN) approach for classifying thyroid nodules (TN) based on historical ultrasound reports. This hierarchical RNN is composed of three layers, with each layer incorporating an individually trained Long Short-Term Memory (LSTM) network. The study's findings indicate that the hierarchical RNN model surpasses basic models in terms of computational efficiency, control accuracy, and robustness, making it an effective tool for diagnosing TN. These advantageous attributes stem from the inherent memory mechanisms of RNNs, which allow them to remember previous states through feedback loops. This memory capability renders RNNs a popular choice for applications in cancer detection.

\vskip2mm

\noindent \textbf{D4. Restricted Boltzmann Machine (RBM):}
A Restricted Boltzmann Machine (RBM) is a type of artificial neural network and a generative stochastic model. It was first introduced by Paul Smolensky \cite{smolensky1986information} in 1986 under the name "Harmonium," but the concept of a "restricted" Boltzmann Machine was developed by Geoffrey Hinton and his students in the mid-2000s.
RBMs have a layer of visible units and a layer of hidden units, but no connections within layers - this is the restriction in their name. Each node in the layer is connected to every node in the other layer. The lack of intra-layer connections simplifies the learning process.
The work by Vairale et al. \cite{vairalephysical} presents an application of Restricted Boltzmann Machines (RBMs) to develop a personalized fitness recommendation system tailored for individuals diagnosed with thyroid conditions. RBMs are a particular class of generative artificial neural networks characterized by a bi-directional architecture, which operates in an unsupervised manner. This structure comprises a visible layer containing binary variables and a hidden layer, also populated with interconnected binary variables. The learning process within RBMs is primarily conducted through statistical analysis.
\vskip2mm

\noindent \textbf{D5. Generative adversarial network (GAN):} This type of ML network is composed of two distinct models: a generator and a discriminator. The generator maps a random input vector to an output in the data space, while the discriminator serves as a binary classifier that evaluates both input data from the training set and output data from the generator. The GAN has gained widespread use in the diagnosis of diseases, including thyroid nodules (TN) \cite{yang2019dscgans, yoo2020generative}.

\subsubsection{Artificial Neural networks (ANN)}
ANNs are defined as a class of information processing systems comprised of interconnected non-linear elements known as neurons. These networks have proven to be effective in addressing complex issues, as they have the ability to store and retrieve information. The various types of ANNs can be divided into various classifications.

\vskip2mm

\noindent \textbf{A1. Extreme learning machine (ELM):}
The ELM model features a layer of hidden nodes with randomized weight distribution. The weights between the hidden node inputs and outputs are learned in a single step, resulting in a more efficient learning process compared to other models. The ELM has been proven to be an effective method in the diagnosis of thyroid disease (TD), as evidenced in several studies such as \cite{li2012computer}, \cite{ma2018efficient}, \cite{xia2017ultrasound} and \cite{pavithra2021optimal}.

\vskip2mm

\noindent \textbf{A2. Multilayer perceptron (MLP):}
MLP represents a category of feedforward networks where data is processed from the input layer through to the output layer. Each layer in this network comprises a varying number of neurons. Rao et al. \cite{rao2019thyroid} introduced an innovative approach for thyroid nodule classification, utilizing MLP with a backpropagation learning algorithm. In their model, the MLP included four neurons in the input layer, three neurons in each of the ten hidden layers, and a single neuron in the output layer.
Hosseinzadeh et al. \cite{hosseinzadeh2020multiple} conducted a separate study with the objective of improving the accuracy of Thyroid Disease (TD) diagnosis through MLP networks. The research compared their findings with existing literature on thyroid cancer classification and found MLP networks to be superior.
Isa et al. \cite{isa2010suitable} delved into the exploration of activation functions within MLP networks. Their goal was to identify the optimal activation function for accurate classification of incurable diseases such as TD and breast cancer. The study evaluated multiple activation functions, including logarithmic, sigmoid, neural, sinusoidal, hyperbolic tangent, and exponential functions. The research found the neural function to be the most effective for TD classification, using the Back Propagation algorithm as the training algorithm. This result was further corroborated by Mourad et al. \cite{mourad2020machine}.

\vskip2mm

\noindent \textbf{A3. Radial basis function (RBF):}
In \cite{erol2008radial}, ML is applied to the classification of TN, where the MLP and RBF activation functions are utilized. The RBF activation function is found to outperform the MLP in terms of the structural classification of thyroid nodules. This approach highlights the effectiveness of activation functions in approximating functions, classifying, and predicting time series data, especially in the diagnosis of thyroid cancer.

\subsubsection{Traditional classification (TCL)}
\vskip2mm
\noindent \textbf{T1. k-nearest neighbors (KNN):}
The nearest k-neighbor (KNN) algorithm is a type of non-parametric supervised machine learning method used for regression and classification. The method relies on the utilization of k-training samples for predictions. In a study conducted by Chandel et al. in \cite{chandel2016comparative}, the KNN method was applied to classify thyroid disease based on TSH, T4, and goiter parameters. Liu et al. \cite{liu2012design} also employed the Fuzzy K-nearest Neighborhood (FKNN) approach to differentiate between hyperthyroidism, hypothyroidism, and normal cases. There is a growing interest in larger datasets for future research, as noted in \cite{geetha2016empirical}.
\vskip2mm

\noindent \textbf{T2. Support vector machines (SVM):}
The Support Vector Machine (SVM) is a machine learning method used for classification and regression tasks. In a study published in \cite{ma2010differential}, an SVM approach was proposed for differentiating benign from malignant thyroid nodules (TN) by utilizing 98 TN samples (82 benign and 16 malignant). Another study in \cite{chang2010application} employed six SVMs to classify nodular thyroid lesions by selecting the most important textural characteristics. The authors reported that the proposed method achieved the correct classification. In \cite{dogantekin2011expert}, a Generalized Discriminant Analysis and Wavelet Carrier Vector Machine system (GDA-WSVM) was introduced for diagnosing TN, consisting of feature extraction, classification, and testing phases.

\vskip2mm

\noindent \textbf{T3. Decision trees (DT):} 
DT learning is a method for data mining that uses a predictive model for decision-making, where the output values are represented by the leaves and the input variables are represented by branches. This approach has been applied to uncover underlying thyroid diseases as demonstrated in various studies such as \cite{yadav2020prediction}, \cite{hao2018prognosis}, \cite{dharmarajan2020thyroid}, and \cite{yadav2019decision}.

\vskip2mm

\noindent \textbf{T4. Logistic regression (LR):} 
In \cite{zhao2015logistic}, the Logistic Regression (LR) model was used to determine the specific characteristics of thyroid microcarcinoma (TMC) in 63 patients, based on the combination of contrast-enhanced ultrasound (CEUS) and conventional US values. Another study, conducted in northern Iran and reported in \cite{yazdani2018factors}, applied LR to analyze 33530 cases of thyroid cancer. LR is a widely used binomial regression model in machine learning.

\subsubsection{Probabilistic models (PM)}

\vskip2mm

\noindent \textbf{P1. Bayesian networks (BN):} 
In computer science and statistics, a Bayesian Network (BN) is a type of model that represents a set of random variables. It has been used to study various diseases, as shown in the references \cite{liu2011bayesian}, \cite{liu2009controlled}, and \cite{ashraf2012hybrid}.

\subsection{Unsupervised learning (USL)}
In AI and computer science, unsupervised learning involves analyzing data without pre-existing labels or annotations. It aims to uncover the underlying structures in the unlabeled data. Unlike supervised learning, which uses labeled data to calculate a success score, unsupervised learning lacks this labeling, making it difficult to assess the accuracy of the results. While unsupervised learning algorithms can perform more complex tasks compared to supervised ones, they can also be more unpredictable, adding unintended categories and introducing noise instead of structure. Despite these challenges, unsupervised learning remains a valuable tool for exploring AI, as it enables the discovery of patterns and relationships in data that might not be immediately apparent \cite{kate2023check, nobile2023unsupervised}.

\subsubsection{Clustering (C)}
The purpose of this method is to segment a set of thyroid cancer data into various homogeneous groups that possess similar characteristics, making it easier to classify the unlabeled datasets into benign and malignant. This detection approach has gained significant attention in various medical studies for its simplicity, including in the detection of DNA copy number changes \cite{manogaran2018machine}, breast cancer recognition \cite{agrawal2019combining}, cancer gene detection \cite{de2008clustering}, skin cancer diagnosis \cite{anas2017skin}, and brain tumor detection \cite{khan2021brain}. Additionally, clustering can also help identify cancer without precise definitions \cite{yu2017clustering}. The clustering technique was used in \cite{chandel2020analysing} to identify factors that impact the normal functioning of TG, and DBSCAN and PCA were applied to manage the clusters and reduce dimensionality. An automated clustering system for thyroid diagnosis was developed in \cite{katikireddy2020performa} to prescribe the appropriate drug datasets for hyperthyroid, hypothyroid, and normal cases. The efficiency of fuzzy clustering for thyroid and liver datasets from the UCI repository was analyzed in \cite{venkataramana2018comparative}, where the FPCM and PFCM algorithms were applied and compared.

\vskip2mm

\noindent \textbf{C1. K-means (KM):} kM method is a technique for data partitioning and a combinatorial optimization challenge. It is commonly utilized in unsupervised learning, in which observations are separated into k groups. In \cite{mahurkar2017normalization}, the authors explore the utilization of Artificial Neural Networks (ANN) and improvised k-Means for normalizing raw data. The study used thyroid data from the UCI dataset containing 215 instances.

\vskip2mm

\noindent \textbf{C2. Entropy-based (EB):} In \cite{yang2019information}, a parameter-free calculation framework named DeMine was developed to predict MRMs. DeMine is a three-step method based on information entropy. Firstly, the miRNA regulation network is transformed into a synergistic miRNA-miRNA network. Then, miRNA clusters are detected by maximizing the entropy density of the target cluster. Finally, the co-regulated mRNAs are integrated into the corresponding clusters to form the final MRMs. The proposed method not only provides improved accuracy but also identifies more miRNAs as potential tumor markers for tumor diagnosis.

\subsubsection{Dimensionality reduction (DR)}
DR is a machine-learning method that transforms data from a high-dimensional space into a lower-dimensional space. This technique is popular for classification due to its cost-effectiveness and ability to eliminate unnecessary data patterns and minimize redundancy. For instance, DR was used to diagnose Thyroid Disease (TD) using cytological images \cite{tarkov2012data}.

\vskip2mm

\textbf{R1. Principal component analysis (PCA):}
PCA is a multivariate statistical method that transforms variables into a reduced set of uncorrelated variables. This approach reduces the number of variables and minimizes redundant information while preserving the relationships between the data as much as possible. PCA has been widely used in cancer detection and classification of benign and malignant thyroid cells. For example, in \cite{shankarlal2020performance}, PCA was utilized to select the optimal set of wavelet coefficients from the application of Double-Tree Complex Wavelet Transform (DTCW) on noisy thyroid images, which were then classified using Random Forest (RF). In \cite{soulaymani2018epidemiological}, PCA was applied to data from 399 patients with three types of thyroid carcinoma (papillary, follicular, and undifferentiated) in Morocco, enabling classification based on factors such as sex, age, type of carcinoma, and region.

\subsection{Ensemble methods (EM)} 
To address the complexity of cancer data and achieve higher accuracy in detection, the use of ensemble methods is commonly employed in the field. This method involves dividing the data into subgroups and applying multiple machine learning techniques to each subgroup simultaneously, then synthesizing the results to make a final diagnosis. By combining multiple models, the ensemble method aims to produce an optimal predictive model for thyroid cancer detection. This approach has been shown to be effective in various studies, such as \cite{chandran2021diagnosis}, where the authors emphasize the importance of ensemble methods in achieving a more comprehensive understanding of the data and improving the accuracy of the diagnosis.

\subsubsection{Bagging (B)}
In the realm of thyroid cancer screening, Bagging is an ensemble learning technique utilized to improve the accuracy and stability of ML algorithms. This algorithm operates by reducing variance and avoiding overfitting and can be applied to a variety of methods, particularly decision trees. The purpose of Bagging is to enhance the performance of weak classifiers in the field of thyroid cancer screening applications.

\vskip2mm

\textbf{B1. Bootstrap aggregation (BA):}
The Bootstrap Aggregating technique is a widely utilized ensemble method aimed at improving the accuracy of Machine Learning algorithms, particularly for the purposes of classification, regression, and variance reduction. In \cite{awujoola975effective}, this approach was employed for diagnosing thyroid abnormalities.

\vskip2mm

\noindent \textbf{B2. Feature bagging (FB):}
In \cite{chen2020diagnosis}, Feature Bagging (FB) is introduced as a method of ensemble learning with the goal of minimizing the correlation between the individual models in the ensemble. FB achieves this by training the models on a randomly selected subset of features, instead of all features in the dataset. The method is applied to distinguish between benign and malignant thyroid cancer cases \cite{himeur2021artificial}.

\subsubsection{Boosting (O)}
Meta-algorithms are often used in unsupervised learning to mitigate the variance and enhance the performance of weak classifiers by transforming them into strong classifiers.

\vskip2mm
\noindent \textbf{O1. Adaboost}
In the study by Pan et al. \cite{pan2016improved}, a new method called AdaBoost was utilized to diagnose thyroid nodules using the standard UCI dataset. The random forest and PCA techniques were employed for classification purposes and to maintain data variability, respectively.

\vskip2mm

\noindent \textbf{O2. Gradient tree boosting (XGBoost)}
In \cite{chen2016xgboost}, the XGBoost algorithm was introduced as a fast and efficient implementation of gradient-boosted decision trees (GTB). Since its introduction, the XGBoost algorithm has been applied to a range of research topics, including civil engineering \cite{lim2019xgboost}, time-series classification \cite{ji2019xg}, sport and health monitoring \cite{guo2019xgboost}, and ischemic stroke readmission \cite{xu2019extreme}.

For thyroid cancer detection, the authors in \cite{chen2020computer} used XGBoost to diagnose benign and malignant thyroid nodules, as a solution to the challenge of obtaining accurate diagnoses with DL models when a large-scale dataset is unavailable.

Table \ref{table:3} provides a summary of research frameworks for the detection of benign and malignant thyroid cancer, including the category, classifier, detected disease, dataset, objective, and used quantifiable metrics. This table helps to categorize AI methods used for thyroid cancer detection and highlights the current key applications. 


\begin{table*}[t!]
\caption{Summary of research frameworks conducted in the detection of thyroid cancer benign and malignant.}
\label{table:3}
\small

\begin{tabular}{
m{1.5cm}
m{1.7cm}
m{1.7cm}
m{1.7cm}
m{1.7cm}
m{1.7cm}
m{2.7cm}
m{2.3cm}
}
\hline

{\small Ref.} & {\small Category} & {\small Classifier} & {\small DD} & {\small Dataset} & {\small O} & {\small SV} & {\small APP} \\ \hline

{\small \cite{ferreira2018autoencoders}} & {\small DL} & {\small DAE} & 
{\small PTC} & {\small TCGA} & {\small O1} & {\small 18985 features} & {\small US} \\ 

{\small \cite{teixeira2017learning}} & {\small DL} & {\small DAE} & {\small PTC} & {\small TCGA} & {\small O1} & {\small 510 samples}  & {\small Omics} \\

{\small \cite{sajeev2020thyroid}} & {\small DL} & {\small CNN} & {\small TC} & {\small PD} & {\small O1} & {\small 10068 images} & {\small Omics} \\

{\small \cite{thomas2020aibx}} & {\small DL} & {\small CNN} & {\small TC} & {\small PD} & {\small O1} & {\small 482 images} & {\small Omics} \\

{\small \cite{kezlarian2020artificial}} & {\small DL} & {\small CNN} & {\small PTC, FTC} & {\small NA} & {\small NA} & {\small NA} & {\small FNAB} \\ 

{\small \cite{sanyal2018artificial}} & {\small DL} & {\small CNN} & {\small PTC} & {\small PD} & {\small O1} & {\small 370 microphotographs} & {\small FNAB} \\ 

{\small \cite{yoon2020artificial}} & {\small DL} & {\small CNN} & {\small PTC} & {\small PD} & {\small O3} & {\small 469 patients} & {\small FNAB} \\ 

{\small \cite{nguyen2020ultrasound}} & {\small DL} & {\small CNN} & {\small TC} & {\small DDTI} & {\small O1} & {\small 298 patients} & {\small US} \\ 

{\small \cite{liu2017classification}} & {\small DL} & {\small CNN} & {\small TC} & {\small  PD} & {\small O1} & {\small 1037 images} & {\small US} \\ 

{\small \cite{abdolali2020automated}} & {\small DL} & {\small CNN} & {\small TN} & {\small PD} & {\small O2} & {\small 80 patients} & {\small US} \\ 

{\small \cite{li2018fully}} & {\small DL} & {\small CNN} & {\small TN} & {\small PD} & {\small O2} & {\small 300 images} & {\small US} \\ 

{\small \cite{kim2016deep}} & {\small DL} & {\small CNN} & {\small TC} & {\small PD} & {\small O1} & {\small 459 labeled} & {\small US} \\ 

{\small \cite{ma2019thyroid}} & {\small DL} & {\small CNN} & {\small TD} & 
{\small ImageNet} & {\small O1} & {\small 2888 samples}  & {\small US }  \\ 

{\small \cite{li2019diagnosis}} & {\small DL} & {\small CNN} & {\small TC} & {\small PD} & {\small O1} & {\small 17627 patients } & {\small US}  \\ 

{\small \cite{xie2019thyroid}} & {\small DL} & {\small CNN} & {\small TC} & 
{\small PD} & {\small O1} & {\small 1110 images } & {\small US } \\ 


{\small \cite{liang2020convolutional}} & {\small DL} & {\small CNN} & 
{\small TN} & {\small PD} & {\small O1, S1} & {\small 537 images} & {\small US}  \\ 

{\small \cite{chen2017thyroid}} & {\small DL} & {\small RNN} & {\small TN} & 
{\small PD} & {\small O1} & {\small 13592 patients}  & {\small US}  \\ 

{\small \cite{vairalephysical}} & {\small DL} & {\small DBM} & {\small TD} & 
{\small PD} & {\small O1} & {\small 94 users } & {\small Fitness}  \\

{\small \cite{yoo2020generative}} & {\small DL} & {\small GAN} & {\small TC}
& {\small PD} & {\small O3} & {\small 109 images} & {\small Surgery} \\ 

{\small \cite{chai2020artificial}} & {\small DL} & {\small NA} & {\small TC} & {\small NA} & {\small NA} & {\small NA} & {\small US} \\ 

{\small \cite{song2019ultrasound}} & {\small DL} & {\small NA} & {\small TC} & {\small PD} & {\small O1} & {\small 1358 images} & {\small US} \\ 

{\small \cite{barczynski2020clinical}} & {\small AI} & {\small NA} & {\small TC} & {\small PD} & {\small O1} & {\small 50 patients} & {\small Surgery} \\ 

{\small \cite{choi2017computer}} & {\small AI} & {\small NA} & {\small TC} & {\small PD} & {\small O1} & {\small 89 patients} & {\small US} \\ 

{\small \cite{li2012computer}} & {\small ANN} & {\small ELM} & {\small TD} & 
{\small UCI} & {\small O1} & {\small 215 patients} & {\small US}  \\

{\small \cite{ma2018efficient}} & {\small ANN} & {\small ELM} & {\small TD}
& {\small UCI} & {\small O1} & {\small 215 patients} & {\small US}  \\ 

{\small \cite{xia2017ultrasound}} & {\small ANN} & {\small ELM} & {\small TD}
& {\small PD} & {\small O1} & {\small 187 patients} & {\small US} \\

{\small \cite{rao2019thyroid}} & {\small ANN} & {\small MLP} & {\small TD} & 
{\small PD} & {\small O1} & {\small 7200 samples}  & {\small US}  \\ 

{\small \cite{hosseinzadeh2020multiple}} & {\small ANN} & {\small MLP} & 
{\small TD} & {\small UCI} & {\small O1} & {\small 7200  patients} & {\small US}  \\ 

{\small \cite{erol2008radial}} & {\small ANN} & {\small RBF} & {\small TD} & 
{\small PD} & {\small O1} & {\small 487 patients} & {\small US}  \\ 

{\small \cite{fragopoulos2020radial}} & {\small ANN} & {\small RBF} & {\small TD} & {\small PD} & {\small O1} & {\small 447 patients} & {\small Cytopathological} \\

{\small \cite{savala2018artificial}} & {\small ANN} & {\small NA} & FTC & {\small PD} & {\small O1} & {\small 57 smears} & {\small FNAB} \\ 

{\small \cite{li2021artificial}} & {\small ANN} & {\small NA} & {\small FTC} & {\small NA} & {\small NA} & {\small NA} & {\small FNAB} \\ 

{\small \cite{zhao2019assessment}} & {\small ANN} & {\small NA} & {\small TC} & {\small TCGA} & {\small O3} & {\small 482 samples} & {\small Histopathological} \\ 

{\small \cite{wildman2019using}} & {\small ANN} & {\small NA} & {\small TC} & {\small PD} & {\small O1} & {\small 1264 patients} & {\small FNAB} \\

{\small \cite{wang2019automatic}} & {\small ANN} & {\small NA} & {\small TN} & {\small PD} & {\small O1} & {\small 276 patients} & {\small US} \\ 

{\small \cite{chandel2016comparative}} & {\small TCL} & {\small KNN} & 
{\small TD} & {\small PD} & {\small O1} & {\small 7200 instances} & {\small US}\\ 

{\small \cite{ozolek2014accurate}} & {\small TCL} & {\small KNN} & FTC & {\small PD} & {\small O1, O2} & {\small 94 patients} & {\small Histopathological} \\ 

{\small \cite{wang2010detection}} & {\small TCL} & {\small SVM} & {\small FTC} & {\small PD} & {\small O1} & {\small 43 nuclei} & {\small Histopathological} \\

{\small \cite{zhu2019deep}} & {\small TCL} & {\small SVM} & TN & {\small PD} & {\small O1} & {\small 467 TN} & {\small US} \\ 

{\small \cite{ma2010differential}} & {\small TCL} & {\small SVM} & {\small TC
} & {\small PD} & {\small O1} & {\small 92 subjects} & {\small US}  \\ 

{\small \cite{bhalla2020expression}} & {\small TCL} & {\small SVM} & {\small PTC} & {\small TCGA} & {\small O1} & {\small 500 patients} & {\small Omics} \\

{\small \cite{dolezal2020deep}} & {\small DL} & {\small DL} & {\small PTC} & {\small TCGA} & {\small O3} & {\small 115 slides} & {\small Omics} \\ 

{\small \cite{daniels2020machine}} & {\small ML} & {\small ML} & {\small TN} & {\small PD} & {\small O1} & {\small 121 patients} & {\small Omics} \\ 

{\small \cite{yadav2020prediction}} & {\small TCL} & {\small DT} & {\small TC%
} & {\small UCI} & {\small O1} & {\small 3739 patients} & {\small US}  \\ 

{\small \cite{dharmarajan2020thyroid}} & {\small TCL} & {\small DT} & 
{\small TC} & {\small NA} & {\small O1} & {\small NA} & {\small US}  \\ 

{\small \cite{yadav2019decision}} & {\small TCL} & {\small DT} & {\small TC}
& {\small UCI} & {\small O1} & {\small 499 patients} & {\small US}  \\ 

{\small \cite{zhao2015logistic}} & {\small TCL} & {\small LR} & {\small TC}
& {\small PD} & {\small O1} & {\small 63 patients} & {\small US}  \\

{\small \cite{yazdani2018factors}} & {\small TCL} & {\small LR} & {\small TN}
& {\small PD} & {\small O1} & {\small 33,530 patients} & {\small US}  \\

{\small \cite{liu2011bayesian}} & {\small PM} & {\small BN} & {\small TD} & 
{\small UCI} & {\small O1} & {\small 93 adult patients} & {\small US}  \\ 

{\small \cite{liu2009controlled}} & {\small PM} & {\small BN} & {\small TC}
& {\small NA} & {\small O1} & {\small 37 patients} & {\small US}  \\ 

{\small \cite{mahurkar2017normalization}} & {\small C} & {\small KM} & 
{\small TC} & {\small UCI} & {\small O1} & {\small 215  instances}  & {\small US}  \\ 

{\small \cite{yang2019information}} & {\small C} & {\small EB} & {\small TC}
& {\small Private data} & {\small O1} & {\small 734 cases} & {\small US} \\ 

{\small \cite{soulaymani2018epidemiological}} & {\small DR} & {\small PCA} & 
{\small TC} & {\small PD} & {\small O1} & {\small NA} & {\small NA}  \\ 

{\small \cite{awujoola975effective}} & {\small B} & {\small BA} & {\small TD}
& {\small WEKA} & {\small O3} & {\small NA} & {\small US} \\ 

{\small \cite{chen2020diagnosis}} & {\small B} & {\small FB} & {\small TN} & 
{\small PD} & {\small O1} & {\small 1480 patients} & {\small US}  \\ 

\hline
\end{tabular}

\end{table*}

\section{Thyroid cancer datasets}
In the field of thyroid carcinoma research, a number of datasets have been developed to facilitate the validation of ML algorithms and models. This is especially important because the creation of such datasets is a major challenge in the area of endocrine ML. In this section, we present an overview of the most significant thyroid databases, which offer a set of standards for evaluating the performance of learning methods and assist in the diagnosis and monitoring of complicated diseases.

\begin{figure*}[t!]
\includegraphics[width=1.8\columnwidth]{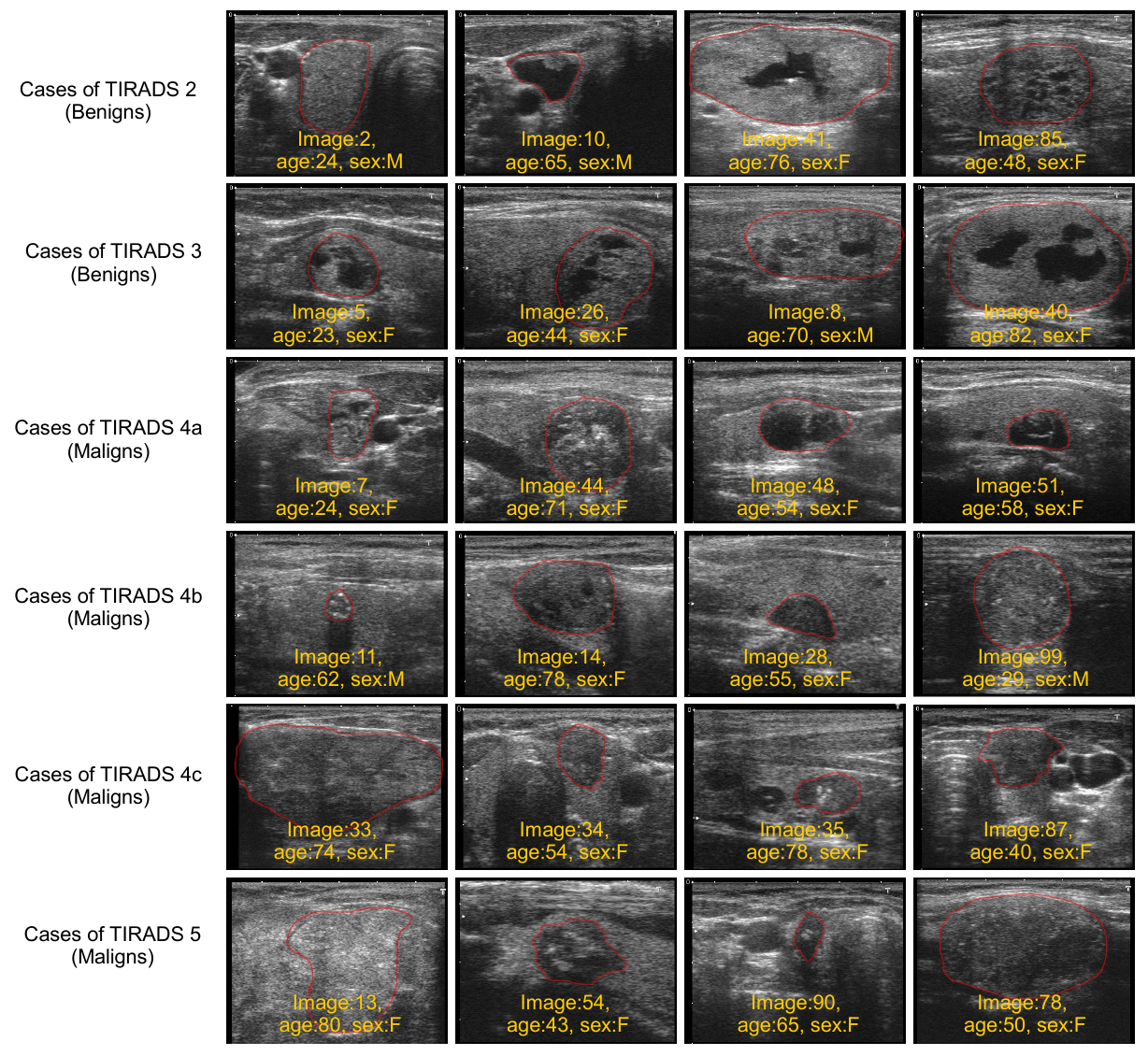}
\caption{Example of six samples for each class from the DDTI datasets.}
\label{fig5}
\end{figure*}

\begin{itemize}
\item \textbf{Waikato Environment for Knowledge Analysis (WEKA):}
The WEKA software, which was created at the University of Waikato using JAVA, is an open-source tool intended for pattern recognition and data analysis tasks such as preprocessing, classification, clustering, correlation, regression, feature selection, and data visualization.

\item \textbf{ThyroidOmics:} 
This is a dataset developed by the Thyroid Working Group of the CHARGE Consortium that aims to examine the underlying factors and consequences of TD using various omics techniques such as genomics, epigenomics, transcriptomics, proteomics, and metabolomics. The dataset consists of the results of the discovery stage of the genome-wide association analysis (GWAS) meta-analysis for thyrotropin (TSH), free thyroxine (FT4), increased TSH (hypothyroidism), and decreased TSH (hyperthyroidism) as reported in \cite{teumer2018genome} and \cite{d1}.

\item \textbf{Thyroid Disease Data Set (TDDS):} 
The dataset utilized for classifying using artificial neural networks (ANN) is referred to as the (dataset name) and features 3772 training instances and 3428 testing instances, with a combination of 15 categorical and 6 real attributes. The three defined classes in this dataset include normal (not hypothyroid), hyperfunction, and subnormal functioning \cite{d2}.

\item \textbf{KEEL Thyroid Dataset:} 
The KEEL dataset provides a set of benchmarks to evaluate the effectiveness of various learning methods. This dataset includes several types of classification, such as standard, multi-instance, imbalanced data, semi-supervised classification, regression, time series, and unsupervised learning, which can be used as reference points for performance analysis \cite{d3}.

\item \textbf{Carcinoma of the thyroid (TNM8):} 
A dataset was created for the purpose of reporting pathologies of thyroid resection specimens associated with carcinoma. The data does not include core needle biopsy specimens or metastasis to the thyroid gland. The dataset also does not encompass NIFTP (Non-invasive Follicular Thyroid Neoplasm with Papillary-like Nuclear Features), tumors of uncertain malignancy (UMP), thyroid carcinomas originating from struma ovarii, carcinomas originating in thyroglossal duct cysts, sarcomas, or lymphomas.

\item \textbf{Gene Expression Omnibus (GEO):} 
The GEO database is a genomics repository that follows the guidelines of the Minimum Information About a Microarray Experiment (MIAME). This database is designed to store gene expression datasets, arrays, and sequences, and provides researchers with access to a vast collection of experiment results, gene profiles, and platform records in GEO \cite{d4}.

\item \textbf{The Surveillance, epidemiology, and end results database (SEER):} 
The creators of this dataset aim to supply a collection of clinical characteristics from thyroid carcinoma patients, which includes 34 details such as age, gender, lymph nodes, etc.

\item \textbf{Digital Database Thyroind Image (DDTI):} 
The DDTI dataset was developed with the support of Universidad Nacional de Colombia, CIM@LAB, and Instituto de Diagnostico Medico (IDIME). It serves as a valuable resource for researchers and new radiologists looking to develop algorithm-based computer-aided diagnosis systems for thyroid nodule analysis. The dataset comprises 99 cases and 134 images, with each patient's data stored in an XML file format \cite{d5}. Fig. \ref{fig5} provides an illustration of six samples from each of the thyroid carcinoma tissue types in the DDTI dataset.

\item \textbf{The Cancer Genome Atlas (TCGA) data:} 
The TCGA is a comprehensive collection of data gathered from 11,000 patients diagnosed with various types of cancer over a period of 12 years. The data consists of detailed genomic, epigenomic, transcriptomic, and proteomic information, amounting to a total of 2.5 petabytes. This extensive dataset has been instrumental in advancing the research, diagnosis, and treatment of cancer.

\item \textbf{The National Cancer Data Repository (NCDR):} 
The NCDR serves as a resource for healthcare and research with the goal of capturing all recorded cases of cancer in England. This data is sourced from the Office for National Statistics \cite{d6}.

\item \textbf{Prostate, Lung, Colorectal, and Ovarian (PLCO) dataset:} 
The National Cancer Institute (NCI) supports the Prostate, Lung, Colorectal, and Ovarian (PLCO) Cancer Screening Trial, aimed at examining the direct factors that contribute to cancer in both men and women. The trial has records of 155,000 participants, and all studies regarding thyroid cancer incidence and mortality can be found within it \cite{d7}.

\end{itemize}

In Table \ref{table:5}, we present examples of public and private thyroid cancer datasets used in thyroid cancer detection.



\begin{table*}[t!]
\caption{Examples of public and private thyroid cancer datasets used in thyroid cancer detection.}
\label{table:5}
\small

\begin{tabular}{
m{12mm}
m{12mm}
m{30mm}
m{20mm}
m{20mm}
m{20mm}
m{20mm}
m{10mm}
}
\hline

{\small Ref} & {\small Year} & {\small TCD} & {\small IT} & {\small IF} & {\small Instance} & {\small M/F} & {\small DA} \\ \hline

{\small \cite{sollini2018texture}} & {\small 2018} & {\small BMU} & {\small Sonographic} & {\small PNG} & {\small 1077} & {\small 4309} & {\small Public} \\

{\small \cite{wang2019automatic}} & {\small 2019} & {\small TCCC} & {\small US} & {\small PNG} & {\small 370} & {\small 370} & {\small Public} \\

{\small \cite{ouyang2019comparison}} & {\small 2019} & {\small Clinical} & {\small US} & {\small JPEG} & {\small 117} & {\small 2108} & {\small Public} \\

{\small \cite{moon2019digital}} & {\small 2019} & {\small Hospital} & {\small US} & {\small JPEG} & {\small 62} & {\small 12/60} & {\small Public} \\

{\small \cite{sun2020evaluation}} & {\small 2020} & {\small Tirads} & {\small US} & {\small JPEG} & {\small 5278} & {\small NA} & {\small Public} \\

{\small \cite{wang2018method}} & {\small 2018} & {\small Peking Union} & {\small US} & {\small JPEG} & {\small 4309} & {\small 1179} & {\small Private} \\

{\small \cite{li2019diagnosis}} & {\small 2019} & {\small Medical Center} & {\small US} & {\small PNG} & {\small 1425} & {\small 2064} & {\small Private} \\

{\small \cite{xu2020computer}} & {\small 2020} & {\small PubMed} & {\small CT scans} & {\small JPEG} & {\small 2108} & {\small 54/253} & {\small Private} \\

{\small \cite{yoon2020artificial}} & {\small 2021} & {\small ACR} & {\small DICOM} & {\small DICOM} & {\small 1629} & {\small 83/289} & {\small Private} \\

{\small \cite{han2021computer}} & {\small 2021} & {\small Clinical} & {\small US} & {\small PNG} & {\small 40} & {\small 407} & {\small Private} \\

\\ \hline
\end{tabular}%
\par
\begin{tablenotes}
\item Abbreviations: Thyroid cancer dataset (TCD); Image Types (IT); Image format (IF); Dataset access (DA); Male (M); Female (F).
\end{tablenotes}

\end{table*}

\textcolor{black}{The trengths and weaknesses of thyroid cancer detection techniques AI-based are summarized in Table \ref{table:6}.}

\begin{table*}[t!]
\caption{A summary of thyroid cancer detection techniques AI-based, including their strengths and weaknesses.}
\label{table:6}
\small

\begin{tabular}{
m{5mm}
m{14mm}
m{60mm}
m{85mm}
}
\hline

{\small Ref.} & {\small AI methods} & {\small Advantages} & {\small Drawbacks} \\ \hline

{\small \cite{li2020classification}} & {\small DAE} & {\small No need labels for thyroid cancer} & {\small Insufficient training data and need relevant data} \\

{\small \cite{ko2019deep}} & {\small CNN} & {\small High thyroid cancer detection} & {\small Insufficient of labels for thyroid cancer and weak in interpretability} \\ 

{\small \cite{lee2019automatic}} & {\small RNN} & {\small No need labels for thyroid cancer} & {\small Slow computation and difficulty in training}  \\ 

{\small \cite{sharifi2021comparison}} & {\small MLP} & {\small Adaptive learning for thyroid cancer}  & {\small Limited in its results} \\ 

{\small \cite{razia2020machine}} & {\small RBF } & {\small Faster training}  & {\small Slow classification} \\ 

{\small \cite{vairale2021recommendation}} & {\small KNN} & {\small High sensitivity to thyroid cancer detection}  & {\small Insufficient of labels for thyroid cancer} \\ 

{\small \cite{shen2020diagnosis}} & {\small SVM} & {\small High sensitivity to thyroid cancer detection} &  {\small Weak in interpretability and long training time} \\ 

{\small \cite{wu2020machine}} & {\small DT} & {\small No require scaling and normalization of data} & {\small Unstable} \\ 

{\small \cite{mcdow2020factors}} & {\small LR} & {\small Low-cost training and easier implementation} & {\small Difficulty to labels data} \\ 

{\small \cite{hsu2019papillary}} & {\small PCA} & {\small Low noise sensitivity} & {\small Loss in information} \\ 

{\small \cite{li2020arb}} & {\small B} & {\small High detection of thyroid cancer} & {\small Loss of interpretability and high computational cost} \\ \hline
\end{tabular}

\end{table*}

\section{Features}
In this section, the focus is on showcasing the crucial techniques utilized in the classification process for characteristic extraction and selection. This primarily involves identifying a subset of relevant features that positively impact the classification accuracy, and eliminating irrelevant variables.

\subsection{Feature selection methods (FS)}

\noindent \textbf{FS1. Information Gain (IG):}
In this section, the focus is on showcasing the most widely-used techniques in the process of classification, with the main objective being to identify and select relevant characteristics that can positively impact the accuracy of classification while eliminating unimportant variables.

Information Gain (IG) is a straightforward method for classifying thyroid cancer features. This method evaluates the likelihood of having cancer by comparing the entropy before and after the examination. Typically, a higher gain value corresponds to lower entropy. IG has been used extensively in several applications for the diagnosis of cancerous diseases, such as in filtering informative genes for precise cancer classification \cite{gao2017hybrid}, selecting breast cancer treatment factors based on the entropy formula \cite{fahrudin2017data}, analyzing and classifying medical data of breast cancer \cite{hamsagayathri2017performance}, reducing the dimensionality of genes in multi-class cancer microarray gene expression datasets \cite{chinnaswamy2017hybrid}, and filtering irrelevant and redundant genes of cancer \cite{gao2017hybrid}. In \cite{mishra2020performance}, IG is utilized as a feature selection technique to eliminate redundant and irrelevant symptoms in datasets related to diabetes, breast cancer, and heart disease. Additionally, the IG-SVM approach, combining Information Gain and Support Vector Machine, has been employed and its results served as input for the LIBSVM classifier \cite{gao2017hybrid}.

\noindent \textbf{FS2. Correlation-based feature selection (CFS):}
The CFS is a technique frequently used for evaluating the correlation between different cancer features. In various studies, the CFS algorithm has been integrated into attribute selection methods for improved classification, such as in \cite{ashraf2013feature} where it was applied to thyroid, hepatitis, and breast cancer data from the UCI ML repository. In \cite{ashraf2012hybrid}, the authors proposed a hybrid method that combined learning algorithm tools and feature selection techniques for disease diagnosis. The CFS was utilized in \cite{al2019gene} for feature selection in microarray datasets to minimize the data's dimensionality and identify discriminatory genes. A hybrid model incorporating the CFS and Binary Particle Swarm Optimization (BPSO) was proposed in \cite{jain2018correlation} to classify cancer types and was applied to 11 benchmark microarray datasets. The CSVM-RFE, which involves the CFS, was used in \cite{rustam2018correlated} to reduce the number of cancer features and eliminate irrelevant ones. In \cite{bhalla2020expression}, the authors utilized CFS-based feature selection techniques to identify key RNA expression features.

\noindent \textbf{FS3. Relief (R):}
The Relief algorithm, commonly known as RA, is an effective method used in selecting important features by assessing their differentiation quality by assigning scores. This technique calculates the weight of various features based on the correlation between cancer attributes. In a study published in \cite{cui2018ovarian}, a feature selection method based on the Relief algorithm was proposed as a means of improving efficiency.

\noindent \textbf{FS4. Consistency-Based Subset Evaluation (CSE):}
The study in \cite{onan2015fuzzy} presents a hybrid classification model for breast cancer, which is based on dividing cancer data into single-class subsets. The effectiveness of the model is evaluated using the Wisconsin Breast Cancer Dataset (WBCD).

\subsection{Feature extraction methods (FE)}

\noindent \textbf{FE1. Principal Components Analysis (PCA):}
The use of Principal Component Analysis (PCA) has been highlighted in several studies as a method to reduce the dimensionality of data and uncorrelated the attributes of cancer features. For instance, in \cite{shankarlal2020performance}, PCA was applied to the dual-tree complex wavelet (DTCW) transform to select the optimum features of Thyroid Cancer. In \cite{soulaymani2018epidemiological}, PCA was proposed as a tool for classifying different thyroid cancer subtypes such as papillary, follicular, and undifferentiated. The implementation of PCA and Linear Discriminant Analysis was also explored in \cite{o2019raman} for classifying Raman spectra of different thyroid cancer subtypes. Finally, in \cite{selaru2004unsupervised}, the authors utilized PCA on cDNA microarray data to uncover the biological basis of breast cancers.

\noindent \textbf{FE2. Texture description (TD):}
Texture analysis is a commonly used method for extracting relevant information in the classification, segmentation, and prediction of Thyroid Cancer. There are numerous texture analysis techniques in the literature, including wavelet transform, binary descriptors, and statistical descriptors. The discrete wavelet transform, in particular, has received significant attention for its ability to perfectly decorrelate data. Many studies have utilized wavelets for thyroid cancer detection, such as in \cite{sudarshan2016application}, where wavelet techniques were employed to identify cancer regions in thyroid, breast, ovarian, and prostate tumors. In \cite{haji2019novel}, texture information was used to diagnose TN malignancy through a 2-level 2D wavelet transform. Other works exploring this area can be found in \cite{yu2019transverse} and \cite{nguyen2019artificial}.

\noindent \textbf{FE3. Active contour (AC):}
The active contour, first introduced by Kass and Witkin in 1987, is a dynamic structure primarily used in image processing. There are several approaches for solving the problem of contour segmentation using a deformable curve model, which has seen numerous applications in the field of Thyroid Cancer detection, as demonstrated in \cite{poudel2017active}, \cite{poudel20163d}, and \cite{nugroho2015thyroid}.

\noindent \textbf{FE4. Local binary patterns (LBP):}
The Local Binary Patterns (LBP) are features employed in computer vision to recognize textures or objects in digital images. LBP has been utilized to detect Thyroid Cancer in \cite{yu2019transverse}. The combination of LBP and DL has also been proposed to classify benign and malignant thyroid nodules in \cite{xie2020hybrid} and \cite{mei2017thyroid}.

\noindent \textbf{FE5. Gray-level co-occurrence matrix (GLCM):}
The Gray-Level Co-occurrence Matrix (GLCM) is a matrix that represents the distribution of values of pixels that occur together at a specified offset in an image. In \cite{song2015model}, GLCM is used to extract features to differentiate between different types of Thyroid Cancer. In \cite{dinvcic2020fractal}, the differences between an individual with Hashimoto's thyroiditis-associated papillary thyroid carcinoma and one with Hashimoto's thyroiditis alone were investigated based on GLCM comparison.

\noindent \textbf{FE6. Independent component analysis (ICA)):}
In Independent Component Analysis (ICA), information is gathered into a set of contributing features for the purpose of feature extraction. ICA is utilized to separate multivariate signals into their individual components. In \cite{kalaimani2019analysis}, ICA is used to extract 29 attributes as independent and useful features for classifying data into either hypothyroid or hyperthyroid using a Support Vector Machine (SVM).

\begin{table*}[t!]
\caption{Summary of feature extraction methods based on DL conducted in the diagnosis of thyroid cancer.}
\label{table:7}
\small

\renewcommand{\arraystretch}{1.5} 
\begin{tabular}{
m{5mm}
m{10mm}
m{10mm}
m{14mm}
m{120mm}
}
\hline

{\small Ref.} & {\small Year} & {\small Classifier} & {\small Features} & {\small Contributions}\\ \hline

{\small \cite{ahmad2017thyroid}} & 2017 & KNN & FC/IG & - Avoid data redundancy and reduce computation time. The kNN deals with the missing dataset, and the ANFIS is provided with the resultant data as input. \\

{\small \cite{nugroho2017classification}} & 2017 & SVM & FC/CFS & - Extract the geometric and moment features while some kernels of the SVM classifier classify the extracted features.\\

{\small \cite{mourad2020machine}} & 2020 & CNN & FC/R & - Combine ML and feature selection algorithms (namely, Fisher’s discriminant ratio, Kruskal-Wallis’ analysis, and Relief-F) to analyse the SEER database. \\

{\small \cite{song2022rapid}} & 2022 & CNN & FE/PCA & - The influence of unbalanced serum Raman data on the prediction results was minimized by using an over-sampling algorithm in this study. PCA then reduced the data's dimension before classifying data using RF and the Adaptive Boosting.\\

{\small \cite{acharya2012thyroscreen}} & 2012 & O & FE/TD & - Combine CAD and DWT and texture feature extraction. The AdaBoost classifier uses the extracted features to classify images into benign or malignant thyroid images. \\

{\small \cite{nugroho2021computer}} & 2021 & CNN & FE/AC & - Image enhancement, segmentation and multi-feature extraction, encompassing both geometric and texture features. Each characteristic is then classified using MLP and SVM, resulting in a determination of either benign or malignant. \\

{\small \cite{sun2020evaluation}} & 2020 & SVM & FE/LBP & - Deep features are extracted by CNN and are combined with the hand-crafted feat- ures, including HOG, LBC, and SIFT to create fused features. These fused features are then used for classification by an SVM. \\

{\small \cite{liu2019value}} & 2019 & SVM & FE/GLCM &  - Use a median filter to reduce noises, and delineate the contours before extracting features from thyroid regions, including GLCM texture features. SVM, random forest, and bootstrap aggregating (bagging) are then used to identify the benign and malignant nodules. \\

{\small \cite{kalaimani2019analysis}} & 2019 & SVM & FE/ICA & Multi-kernel-based is used as a classifier to distinguish the thyroid disease.  \\
\hline
\end{tabular}

\end{table*}

\textcolor{black}{A summary of features methods based on DL conducted in the diagnosis of thyroid cancer are illustrated in Table \ref{table:7}.}

\section{Standard assessment criteria}
In this section, we examine the most commonly utilized standard parameters for evaluating the identification of Thyroid Diseases (TD). These criteria serve as a measure of the effectiveness of the methods used. Selecting the right metric is crucial when evaluating the performance of machine learning models. Numerous metrics have been proposed to evaluate machine learning models in various applications. Here, we present a summary of popular metrics that are considered suitable for assessing the performance of AI algorithms applied in the detection of Thyroid Cancer.
\textcolor{black}{(See Table \ref{table:8}, \ref{table:9} and \ref{table:10})}

\subsection{Classification and Regression Metrics}
Table \ref{table:8} presents an outline of classification and regression metrics used in evaluating AI-based thyroid cancer detection frameworks.

\begin{table*}[!t]
\caption{Summary of classification and regression metrics used in evaluating AI-based thyroid cancer detection schemes.}
\label{table:8}
\small
\begin{tabular}{
m{3cm}
m{5.2cm}
m{8.5cm}
}
\hline
Metric & Mathematical formula & Description   \\ \hline
Accuracy (ACC) & $ACC=\frac{T_{P}+T_{N}}{T_{P}+F_{P}+T_{N}+F_{N}}100\%$ & Give the correct percent of the total number of positive and negative predictions. \\

Specificity (SPE)  & $SPE=\frac{T_{N}}{T_{N}+F_{P}}100\%$ & It is the ratio of correctly predicted negative samples to the total negative samples.  \\

Sensitivity (SEN)  & $SEN= \frac{T_{P}}{T_{P}+F_{N}}100\%$  & It is a quantifiable measure metric of real positive cases that got predicted as true positive cases.  \\

Precision (P)  & $P=\frac{T_{P}}{T_{P}+F_{P}}100\%$ & Measure the proportion of true positive predictions made by the model, out of all the positive predictions made by the model.  \\

F1 Score (F1)  & $F1 = 2\times \frac{Precision \times Recall}{Precision + Recall}$  & It is the harmonic mean of precision and sensitivity of the classification.  \\

Error Rate (ER)  & $ER= \frac{F_{N}+F_{P}}{T_{P}+F_{N}+F_{P}+T_{N}}100\%$  & It is equivalent to 1 minus Accuracy.  \\

Root mean square error (RMSE)  & $RMSE=\left( \sqrt{1-\left( ER\right) ^{2}}\right) \times SD$  & It is the standard deviation of the predicted error between the training and testing dataset, its lower value means that the classifier is an excellent one.  \\

The negative predictive value (NPV)  & $NPV=\frac{T_{N}}{T_{N}+F_{N}}$  & is the proportions of negative results in diagnostic tests, its higher value means the accuracy in diagnosis.   \\

Jaccard similarity index (JSI)  & $JSC=\frac{\left\vert A\cap B\right\vert }{\left\vert A\cup B\right\vert }= \frac{T_{P}}{T_{P}+F_{P}+F_{N}}$ & It has been proposed by Paul Jaccard to gauge the similarity and variety in samples.  \\

Fallout or false positive rate (FPR) & $FPR=\frac{F_{P}}{F_{P}+T_{N}}=1-SP$  & Measure the proportion of negative samples that are incorrectly classified as positive by the model.  \\

Volumetric Overlap Error (VOE)  & $VOE=\frac{F_{P}+F_{N}}{T_{P}+F_{P}+F_{N}}$  & Evaluate the similarity between the segmented region and the ground truth region. VOE measures the amount of overlap between the two regions and is defined as the ratio of the volume of the union of the segmented region and the ground truth region to the volume of their intersection.  \\

Mean Absolute Error (MAE)  & $MAE=\frac{1}{N}\sum\limits_{i=1}^{N}\left\vert a_{i}-p_{i}\right\vert$  & It is the average of the difference between the original values and the predicted values.  \\

Mean Squared Error (MSE)  & $MSE=\frac{1}{N}\sum_{i=1}^{N}\left( y_{i}-r_{i}\right) ^{2}$  & It is the average of the square of the difference between the original values and the predicted values.  \\ \hline

\end{tabular}
\end{table*}

\subsection{Statistical  Metrics}
Table \ref{table:9} depicts a summary of statistical metrics used in assessing AI-based thyroid cancer detection schemes.

\begin{table*}[!t]
\caption{Summary of statistical metrics used in assessing AI-based thyroid cancer detection schemes.}
\label{table:9}

\small
\begin{tabular}{
m{3cm}
m{5.2cm}
m{8.5cm}
}
\hline
Metric & Mathematical formula & Description   \\ \hline

Standard deviation (SD)  & $\sigma =\sqrt{\sum \left( x-\mu \right)^{2}/N}$  & It is a measure of the amount of variation or dispersion in a set of data.  \\

Correlation (Corr)  & $r=(\sum ((x-\mu x)\cdot (y-\mu y)))/(\sqrt{(\sum (x-\mu x)^{2})}\cdot \sqrt{(\sum (y-\mu y)^{2}))}$  & It describes the degree of association or relationship between two or more variables.   \\

Kappa de Cohen & $k=\frac{\Pr \left( a\right) -\Pr \left( e\right) }{1-\Pr \left( e\right) }$  & It measures the degree of concordance between two evaluators, relative to chance.   \\ \hline

\end{tabular}
\end{table*}

\subsection{Computer Vision Metrics}
Table \ref{table:10} portrays a summary of computer vision metrics used in assessing AI-based thyroid cancer detection schemes.

\begin{table*}[!t]
\caption{Summary of computer vision metrics used in assessing AI-based thyroid cancer detection schemes.}
\label{table:10}

\small
\begin{tabular}{
m{3cm}
m{5.2cm}
m{8.5cm}
}
\hline
Metric & Mathematical formula & Description   \\ \hline

Peak Signal to Noise Ratio (PSNR)  & $PSNR=10\cdot \log_{10}((MAX_{I}^{2})/MSE)$  & It measures the ratio of the maximum possible power of a signal to the power of the noise that affects the fidelity of its representation.  \\

\multicolumn{3}{c}{} \\

Structural Similarity Index (SSIM) & $MSSIM(x,y)=\frac{1}{L}\sum_{i=1}^{L} SSIM(x_{i},y_{i})$  & It evaluates the similarity between two images or videos by comparing their luminance, contrast, and structural information. \\

\multicolumn{3}{c}{} \\

Visual Information Fidelity (VIF)  & $VIF=\frac{\sum_{j}I(C^{j};F^{j}/s^{j})}{\sum_{j}I(C^{j};E^{j}/s^{j})}$  & It evaluates the quality of a reconstructed or compressed image or video compared to the original signal. It measures the amount of visual information preserved in the processed image or video, taking into account the spatial and frequency characteristics of the image.  \\

\multicolumn{3}{c}{} \\

Normalized Cross-Correlation (NCC)  & $NCC=\frac{\sum_{i=1}^{M}\sum_{i=1}^{N}(I(i,j)-R(i,j))^{2}}{\sum_{i=1}^{M}\sum_{i=1}^{N}I(i,j)^{2}}$  & Measure the similarity between two images (or videos) by subtracting the mean value of each signal from the signal itself. Then, the signals are normalized by dividing them by their standard deviation. Finally, the cross-correlation between the two normalized signals is calculated.  \\

\multicolumn{3}{c}{} \\

Structural Content (SC)  & $SC=\frac{\sum_{i=1}^{M}\sum_{j=1}^{N}I(i,j)^{2}}{\sum_{i=1}^{M}\sum_{j=1}^{N}R(i,j)^{2}}$  & A higher value of SC (Structural Content) shows that the image is of poor quality.  \\

\multicolumn{3}{c}{} \\

Weight Peak Signal to Noise Ratio (WPSNR)  & $WPSNR=10\log \left( \frac{(2^{n}-1)^{2}}{NVF\times MSE}\right)$  & It takes into account the image texture \cite{erfurt2019study}. \\

\multicolumn{3}{c}{} \\

Noise Visibility Function (NVF)  & $NVF=NORMALIZATION\left\{ \frac{1}{1+\delta_{bloc}^{2}}\right\}$ \newline where $\delta_{bloc}$ is the luminance variance.  & It estimates the texture content in the image. \\

\multicolumn{3}{c}{} \\

Visual Signal to Noise Ratio (VSNR)  & $VSNR=10\log_{10}\left( \frac{C^{2}(I)}{(VD)^{2}}\right)$ \newline Where $C(I)$ is the RMS contrast of the original image $I$ and $VD$ is visual distortion \cite{chandler2007vsnr}.  & It is based on the specified thresholds of distortions in the image based on the computing of contrast thresholds and wavelet transform. If the distortions are lower than the threshold, the VSNR is perfect.  \\

\multicolumn{3}{c}{} \\

Weighted signal-to-noise ratio (WSNR)  & $WSNR=10\log_{10}\left( \frac{\sum_{u=0}^{M-1}\sum_{v=0}^{N-1}\left\vert A(u,v) C(u,v) \right\vert ^{2}}{\sum_{u=0}^{M-1}\sum_{v=0}^{N-1}\left\vert A(u,v) -B(u,v) C(u,v) \right\vert ^{2}}\right)$  \newline where $A(u,v)$, $B(u,v)$, and $C(u,v)$ represent the Discrete Fourier Transforms (2D TFD) \cite{zhou2020weighted}. & It is based on the contrast sensitivity function (CSF).  \\

\multicolumn{3}{c}{} \\

Normalized Absolute Error (NAE):  & $NAE=\frac{\sum_{i=1}^{M}\sum_{j=1}^{N}\left\vert I(i,j)-R(i,j) \right\vert }{\sum_{i=1}^{M}\sum_{j=1}^{N}\left\vert I(i,j) \right\vert }$  & It evaluates the accuracy of an ML model's predictions. It measures the difference between the predicted values and the actual values, as a proportion of the range of the actual values.  \\

\multicolumn{3}{c}{} \\

Laplacian Mean Square Error (LMSE)  & $LMSE=\frac{\sum_{i=1}^{M}\sum_{j=1}^{N}\left[ L(I(i,j)) -L(R(i,j)) \right]^{2}}{\sum_{i=1}^{M}\sum_{j=1}^{N}\left[ L(I(i,j)) \right]^{2}}$ \newline where $L(I(i,j))$ is the Laplacian operator.  & It is a variant of the Mean Square Error (MSE) that uses the Laplacian distribution instead of the Gaussian distribution.  \\ \hline

\end{tabular}
\end{table*}

\subsection{Ranking Metrics}
\noindent \textbf{M1. The mean reciprocal rank (MRR):}
The MRR is a statistic measure for evaluating the mean reciprocal rank of results for a sample of queries\cite{lasseck2018audio}.

\begin{equation}
MRR=\frac{1}{\left\vert Q\right\vert }\sum_{i=1}^{\left\vert Q\right\vert }%
\frac{1}{rank_{i}}
\end{equation}

Where rank$_{i}$  refers to the rank position of the first relevant document for the i-th query.\\

\noindent \textbf{M2. The Discounted cumulative gain (DCG):} the DCG is used to measure the ranking quality \cite{agarwal2018counterfactual}.

\section{Example of thyroid cancer detection using AI}
To explain how thyroid cancer has been considered in the literature and how AI can be used to detect types of cancers. In the following, we present a simple example to classify TD. It has been known that pattern recognition is the process of training a neural network to assign the correct target classes to a set of input patterns. Once trained the network can be used to classify patterns. In this section, we present an example of thyroid cancer classification as benign, malignant, and normal based on a set of features specified according to the TIRADS.  In this example, the dataset (7200 samples) is selected from the UCI Machine Learning Repository \cite{murphy1994uci}. This dataset can be used to create a neural network that classifies patients referred to a clinic as normal, hyperfunction, or subnormal functioning. The Thyroid Inputs (TI) and Thyroid Targets (TT) are defined as: (i) TI: a 21x7200 matrix consisting of 7200 patients characterized by 15 binary and 6 continuous patient attributes. (ii) TT: a 3x7200 matrix of 7200 associated class vectors defining which of three classes each input is assigned to. Classes are represented by a 1 in rows 1, 2, or 3.  (1) Normal, not hyperthyroid. (2) Hyperfunction. (3) Subnormal functioning.

\begin{figure*}[t!]
\begin{center}
\includegraphics[width=1.7\columnwidth]{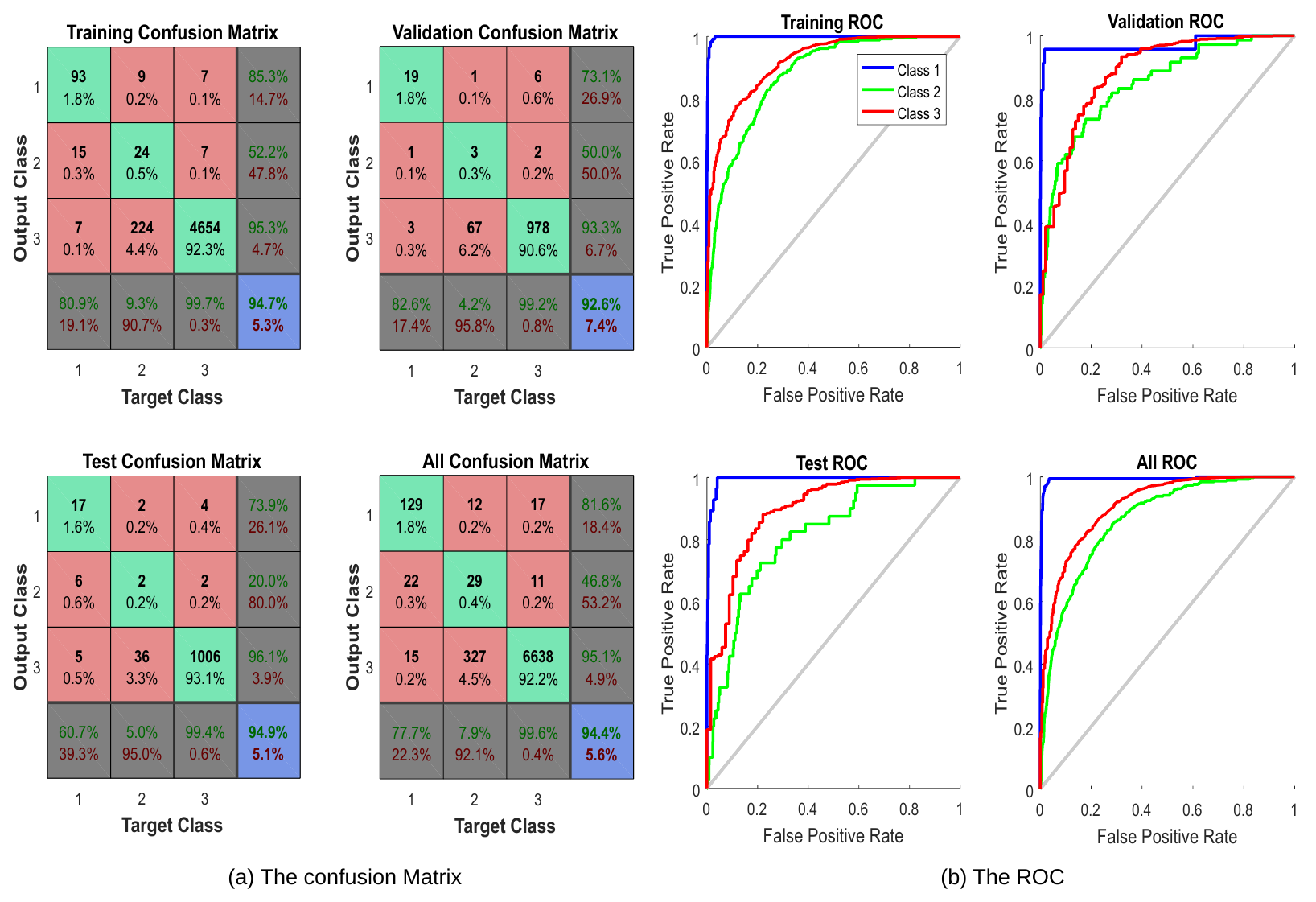}
\end{center}
\caption{ An example of the confusion matrix and ROC metric for thyroid cancer classification.}
\label{fig6}
\end{figure*}

In this network, the data is divided into 5040 samples, 1080 samples, and 1080 samples used for training, validation, and testing respectively. The network is trained to reduce the error between thyroid inputs and thyroid targets or until it reaches the target goal. If the error rate does not decrease and the training does not improve, the training data is halted with data of validation. The data testing is used to deduce the values of targets. Thus, it determines the percentage of learning. For this example, the 10 hidden layer neurons are used in this model for 21 input and 3 output. After the simulation of the model, the Percent Error was been 5.337\%, 7.407\%, and 5.092\% for training, validation, and testing respectively. Thus in the total, it recognized 94.4\% and the overall error rate was 5.6\%. The confusion matrix and the ROC metric are illustrated in Fig.\,\ref{fig6}.

\begin{figure*}[t!]
\begin{center}
\includegraphics[width=1.8\columnwidth]{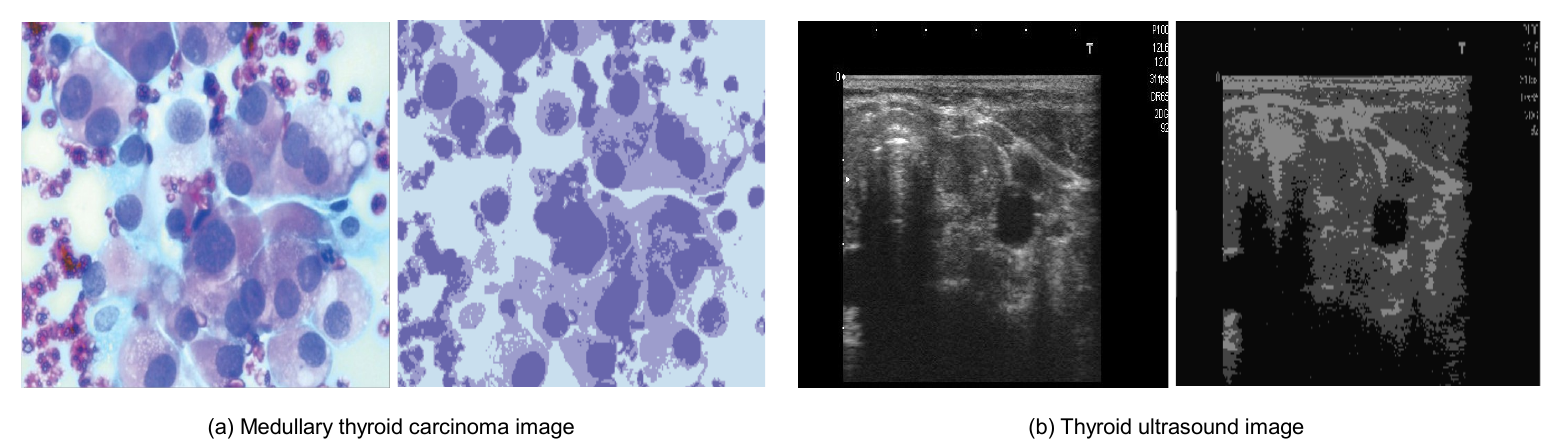}
\end{center}
\caption{ Example of thyroid segmentation based on the k-means method.}
\label{fig7}
\end{figure*}

Fig.\,\ref{fig7}, illustrates an example of thyroid segmentation (TS) in ultrasound images using k-means (3 clusters have been chosen for this example) which is one of the most commonly used clustering techniques. 

\section{Critical analysis and discussion}
As we delve into the core of this paper, it is essential to critically assess and discuss the multitude of facets associated with the application of AI in thyroid carcinoma detection. While the promise of AI has been well-articulated in existing literature, a more nuanced perspective is needed to fully understand its impact on healthcare, both positive and negative.
In this section, we undertake a critical analysis of the effectiveness of AI models for thyroid carcinoma detection. Moving beyond the optimistic numbers, we will question the robustness of these models in real-world clinical settings and discuss their role in the broader context of clinical decision-making.
Furthermore, we explore the potential biases in AI models, understanding how they might inadvertently perpetuate existing inequities in healthcare. A comparative assessment of AI-based and traditional diagnostic methods will provide deeper insights into their relative effectiveness.
Moving on, acknowledging the challenges to the implementation of AI tools in healthcare, we delve into the infrastructural, regulatory, and cultural barriers that might hinder their widespread adoption. Lastly, we underscore the crucial role of interdisciplinary collaboration in ensuring the successful integration of AI into healthcare.

\textcolor{black}{A summary of features methods based on DL conducted in the diagnosis of thyroid cancer are detailed in Table \ref{tab11}.}


The Effectiveness of AI Models: The reported accuracy, sensitivity, and specificity of AI models in the literature may vary widely based on the dataset used, the quality of the data, and the methodology employed. AI models' effectiveness in a controlled experimental environment may not reflect their performance in a real-world clinical setting. Factors like noise in the data, incomplete data, and changing clinical conditions can dramatically influence the outcome. Therefore, it is crucial to scrutinize the model's robustness and reliability under various conditions.

\begin{table*}[t!]
\caption{\textcolor{black}{Performance evaluation of various thyroid cancer frameworks}}
\label{tab11}
\small

\begin{tabular}{
m{6mm}
m{20mm}
m{24mm}
m{12mm}
m{12mm}
m{12mm}
m{12mm}
m{12mm}
m{12mm}
m{12mm}
}
\hline

Ref. &  AI Model & Dataset	& ACC &	SPE	& SEN & PPV & F1 & NPV  & AUC\\ \hline

\cite{bai2020thyroid}& CNN &  PD& 88.00 & 79.10 & 98.10 & - & - & - & - \\ 

 \cite{xia2017ultrasound} & ELM &PD & 87.72 & 94.55 & 78.89 & - & - & - & - \\ 

 \cite{mourad2020machine} & MLP  &PD & 87.16  & 87.05 & 91.18 & 0.162 & 0.275 & 0.997 & - \\ 

 \cite{chandran2021diagnosis} & SVM &PD & 63.27 & 71.85 & 38.46 & 32.43 & - & 76.87 & - \\

 \cite{colakoglu2019diagnostic} & RF &PD & 86.8 & 87.9 & 85.2 & - & - & - & 0.92 \\ 

 \cite{park2021combining} & LR & PD& 77.8 & 79.8 & 70.6 & - & - & - & 0.75 \\ 

 \cite{liu2019value} & B & PD& 84.69 & 0.8696 & 0.8269 & 0.8776 & - & 0.8163 & 0.8852\\ 

\cite{duc2022ensemble} & Ensemble DL & Cytological images & 99.71 & - & - & - & -  & - & - \\

\cite{guan2019deep} & VGG-16 & Cytological images & 97.66 & - & - & - & - & - \\

\cite{lin2021deep} & VGG-16 &  & 99 & 86 & 94 & - & 88 & - & - \\

\cite{zhang2019machine} & RF & Ultrasound  & - & - & - & - & - & - & 0.94 \\

\cite{ouyang2019comparison} & k-SVM & Ultrasound & -& - & - & - & - & - & 0.95 \\

\cite{shin2020application} & ANN & Ultrasound & -& - & - & - & - & - & 0.69 \\

\cite{zhao2021comparative} & SVM RF & Ultrasound & - & - & - & - & - & - & 0.951 \\

\cite{vadhiraj2021ultrasound} & ANN SVM & Ultrasound & 0.96& - & - & - & - & - & - \\

\cite{gild2022risk} & RF & Ultrasound & -& - & - & - & - & - & 0.75 \\

\cite{ma2017pre} & CNN & DICOM & 83.00& 85.00 & 82.40 & - & - & - & - \\

\cite{ma2017ultrasound} & CNN & DICOM & -& 91.50 & - & - & - & - & - \\

\cite{chi2017thyroid} & Fine-Tuning DCNN & PD & 99.10& - & - & - & - & - & - \\

\cite{zhu2017image} & ResNet18-based network &  PD& 93.80& - & - & - & - & - & - \\

\cite{gao2018computer} & multiple-scale CNN & PD & 82.20& - & - & - & - & - & - \\

\cite{peng2021deep} & ThyNet & PD & -& - & - & - & - & - & 0.921 \\

\cite{zuo2018extraction} & Alexnet CNN & PD & 86.00& - & - & - & - & - & - \\

\cite{zhu2019deep} & DNN & ACR TIRADS & 87.20& - & - & - & - & - & - \\

\cite{zhu2021efficient} & CNN (BETNET) & Ultrasound &98.30 & - & - & - & - & - & - \\

\cite{kim2022deep} & ResNet & TIRADS & 75.00& - & - & - & - & - & - \\

\cite{lee2019application} & Xception & CT images & 89.00& 92.00 & 86.00 & - & - & - & - \\

\cite{li2019diagnosis} & DCNN & Sono graphic images & 89.00& 86.00 & 93.00 & - & - & - & - \\

\cite{tsou2019mapping} & Google inception v3 & Histo pathology images & 95.00& - & - & - & - & - & - \\

\cite{lu2020application} & Cascade MaskR-CNN & Ultrasound &94.00 & 95.00 & 93.00 & - & - & - & - \\

\cite{kwon2020ultrasonographic} & VGG16 & Ultrasound & -& 92.00 & 70.00 & - & - & - & - \\

\cite{chan2021using} & VGG19 & Ultrasound &77.60 & 81.40 & 72.50 & - & - & - & - \\

\cite{wang2020comparison} & VGG16 & Ultrasound & 74.00& 80.00 & 63.00 & - & - & - & - \\

\cite{sun2020evaluation} & SVM CNN & Ultrasound  & 92.50& 83.10 & 96.40 & - & - & - & - \\

\cite{kim2021convolutional} & CNN & TIRADS  & 85.10& 86.10 & 81.80 & - & - & - & - \\

\cite{wu2021deep} & CNN & TIRADS  & 82.10& 85.00 & 78.00 & - & - & - & - \\

\cite{jin2020ultrasound} & CNN & TIRADS  & 80.30& 80.10 & 80.60 & - & - & - & - \\

\cite{wang2021radiomic} & CNN & US  & 83.00& 47.00 & 65.00 & - & - & - & - \\

\cite{wei2021radiomics} & CNN & MRI  & 79.00& 80.00 & 65.00 & - & - & - & - \\

\cite{zhou2020differential} & CNN & US  & 97.00& 84.10 & 89.50 & - & - & - & - \\

\cite{gu2019prediction} & CNN & CT image  & 84.00& 73.00 & 93.00 & - & - & - & - \\

\cite{park2019association} & CNN & US  & 77.00& - & - & - & - & - & - \\

\hline

\hline

\end{tabular}

\end{table*}

\subsection{Limitations and open challenges}
Despite the success of AI tools in thyroid cancer diagnosis, their limitations hinder the development of effective solutions, make their application costly, and limit their diffusion. For instance,to achieve precise thyroid cancer detection, it is crucial to gather and store all relevant data in one place. Then, algorithms must be developed to identify all forms of thyroid cancer. Every thyroid cancer dataset includes a set of training images, test images, nodule plans, and classifications of nodule characteristics of diverse sizes \cite{shah2023deep}. The datasets must be regularly updated using MRI, CT scans, X-rays, and clinically obtained scans to assess thyroid conditions, and they should also include demographic information such as gender and age. Additionally, it is important to establish a unified and centralized database accessible to all medical centers to test, validate, and apply AI algorithms to existing data \cite{salazar2019thyroid}. Moving, the rest of the limitations and open challenges can be summarized as follows:


\begin{itemize}[leftmargin=*]
\item Insufficient clean data and accuracy: The lack of comprehensive and annotated data sets regarding the incidence and spread of cancer, specifically thyroid cancer, is a major hindrance to accurate cancer diagnoses and efficient treatment. Medical statistics often do not properly record the number of deaths caused by thyroid cancer, making data collection and validation challenging \cite{elmore2021blueprint}. This results in a limited amount of data typically collected from one center, due to the absence of a dedicated thyroid cancer clinical database shared among institutions. The accuracy of AI algorithms in diagnosing thyroid cancer is also limited by the limited number of available labeled cases for clinical outcomes \cite{park2021key}. Researchers acknowledge that a large amount of data is necessary for the neural network to yield accurate results, but caution must be taken in regard to the data added during the learning phase as it can introduce noise. 

\item Thyroid gland imaging:  In the diagnostic evaluation of thyroid cancer, computed tomography (CT) and magnetic resonance imaging (MRI) are available options but they are not considered the preferred methods due to their high cost and unavailability in certain cases \cite{ha2021applications}. Instead, ultrasound is commonly used as an alternative to physical exams, radioisotope scans, or fine-needle aspiration biopsies. During a ultrasound examination, the doctor is able to assess the activity of the gland by observing the echo of the node and determining its echogenicity, size, limits, and the presence of calcifications. However, the results obtained from ultrasound tests are not always accurate enough to differentiate between benign and malignant nodes and the images obtained can be more prone to noise \cite{zhu2021generic}.

\item The number of DL layers:  Choosing the right DL algorithm is crucial in addressing various issues, particularly those related to thyroid cancer diagnosis. Due to the close similarities between benign and malignant tumors, as well as between tumors and other types of lymphocytes, it is challenging to differentiate between them accurately \cite{wang2022soft}. To achieve this, a significant increase in the number of layers for feature extraction may be required. However, this results in a longer processing time, especially when dealing with large amounts of data, which can impact the timeliness of the diagnosis for cancer patients \cite{lin2021deep}.    

\item The computation cost and space: In the field of algorithms, time computing is a metric that assesses the computational complexity of an algorithm, which predicts the time it will take to run the algorithm by calculating the number of basic operations it performs, as well as its dependence on the size of the input. Typically, time computing is expressed as $O(n)$, where $n$ represents the size of the input, measured in terms of the number of bits required to represent it \cite{al2021cost}. Researchers in the AI field, especially those working on thyroid cancer or other types of cancer diagnosis, face the challenge of finding algorithms that are both highly accurate and efficient in terms of processing time. They aim to develop algorithms that can analyze vast amounts of data quickly while still providing accurate results. Moreover, the volume of data used in these algorithms can sometimes exceed the available storage space \cite{lin2021deep}.

\item Imbalanced dataset: The distribution of cancer elements within categories related to thyroid tissue cells is often uneven, as these cells often make up a minority of the total tissue cell dataset. As a result, the data set is highly imbalanced, consisting of both cancer cells and normal cells. This unbalanced distribution of features in cancer cell detection datasets often results in the suboptimal performance of AI algorithms used for the detection \cite{yao2022deepthy}. 

\item Sparse labels: Labeling is a crucial aspect of CT detection, specifically for distinguishing between normal and abnormal cancer cells. However, the process can be time-consuming and costly due to the limited number of available labels. This scarcity results in inconsistent decisions and can negatively impact the accuracy of AI algorithms, which heavily rely on labeled data. This can eventually undermine the trust and credibility of this type of application \cite{yao2022deepthy}. 

\item The volume of data: At present, with the advancement in technology, especially in the field of thyroid cancer diagnosis and the growing volume of medical and patient data, researchers are facing challenges in suggesting algorithms that can effectively handle a limited number of samples, noisy samples, unannotated samples, sparse samples, incomplete samples, and high-dimensional samples. This requires AI algorithms that are highly efficient and capable of processing vast amounts of data exchanged between healthcare providers and patients or among specialist physicians \cite{sayed2023time}

\item The error-susceptibility: Despite AI being self-sufficient, it is still susceptible to errors. For instance, when training an algorithm with thyroid cancer datasets to diagnose cancerous regions, it can result in biased predictions if the training sets are biased. This can lead to a series of incorrect results that may go unnoticed for an extended period. If detected, identifying and correcting the source of the problem can be a time-consuming process \cite{karsa2020optimized}.

\item The data form: Despite the numerous advancements in the use of AI for thyroid cancer detection, several limitations persist and pose a challenge to its progress. With the growing demand for various medical imaging technologies that result in vast amounts of data needed for AI algorithms, coordinating and organizing this information has become a daunting task. This can largely be attributed to the absence of proper labeling, annotation, or segmentation of the data, making it difficult to manage effectively \cite{kim2022second}.

\item Unexplainable AI: The utilization of AI in the medical field can sometimes yield results that are unclear and lack proper justification, known as a "black box". This leaves doctors unsure about the accuracy of the results and may lead to erroneous decisions and treatments for patients with thyroid cancer. Essentially, AI can behave like a black box and fail to provide understandable explanations for its outputs \cite{sardianos2021emergence}.

\item Lack of cancer detection platform: One of the major barriers to detecting various cancers, particularly thyroid cancer, is the limited availability of platforms for reproducing and examining previous results. This shortage represents a significant weakness and hinders the comparison of AI algorithm performance, making it challenging to improve their efficacy \cite{abdolali2020automated}. The presence of online platforms with comprehensive data sets, cutting-edge algorithms, and expert recommendations is vital in aiding doctors, researchers, developers, and specialists to make informed decisions with a low margin of error. Such platforms also provide a crucial supplement to clinical diagnoses by allowing for more comprehensive experimentation and comparison \cite{masuda2021machine}.

\item The digitization and loss data: Digitization of medical records has become a necessity, particularly in the realm of cancer diagnosis, due to the widespread adoption of various technologies such as Whole Slide Images (WSIs). WSIs, which serve as digital versions of glass slides, facilitate the application of AI techniques for pathological analysis \cite{dov2019thyroid}. Despite its benefits, digitization in the medical field is confronted with certain limitations, such as the risk of significant information loss during quantification and inaccuracies that may arise from data compression utilized in autoencoder algorithms. Hence, it is crucial to be mindful in selecting the right digitization technology to preserve the information and maintain the originality of the data \cite{dovai,halicek2019head}.

\item The Contrast: The absence of sufficient contrast in the tissues neighboring the thyroid gland (TG) complicates the process of accurately analyzing and diagnosing thyroid cancer.

\end{itemize}




\begin{figure*}[t!]
\centering
\includegraphics[width=1.8\columnwidth]{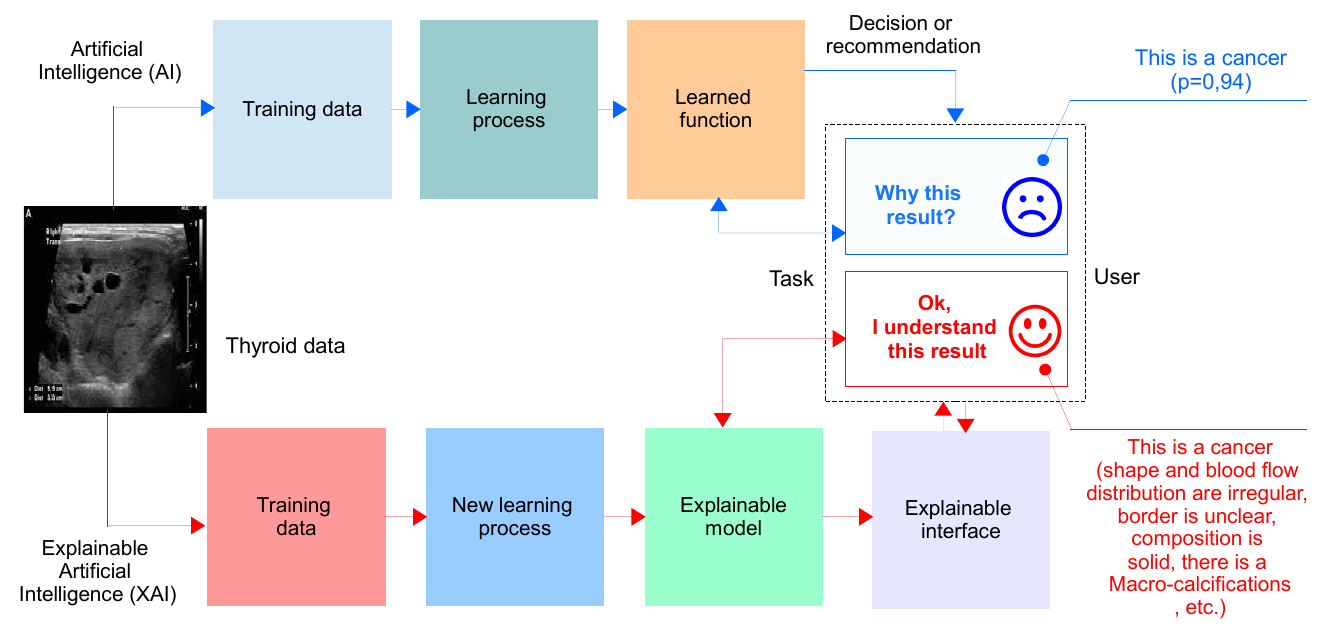}
\caption{Explainable Artificial Intelligence (XAI) diagram block.}
\label{fig8}
\end{figure*}

\section{Future Research directions}
We also highlight the future trajectory of AI in thyroid carcinoma detection, discussing emerging trends and technologies while considering their ethical implications. The ethical considerations do not end there, as we further examine issues related to data privacy, accountability, and equity.
This section highlights promising research trends that will have a major effect on enhancing thyroid cancer detection in the future. 

\subsection{Explainable Artificial Intelligence (XAI)} 
The use of artificial intelligence (AI) systems in decision-making is crucial, but they can be complex and difficult to understand. To address this issue, the field of explainable AI (XAI) has emerged, which aims to provide transparency in AI models. The need for XAI is especially important in health applications where the interpretation of results is crucial. The use of XAI has been demonstrated in the analysis of incurable diseases affecting the TG, as seen in several studies such as \cite{lamy2019explainable, kobylinska2019explainable, pintelas2020explainable, lamy2020intelligence, poceviciute2020survey}. The difference between AI and XAI is illustrated in Fig. \ref{fig8}. In \cite{wildman2019using}, the authors present an XAI model for the detection of thyroid cancer, which improves the confidence of medical practitioners in the predictions. Unlike traditional AI algorithms, XAI models provide evidence to support their conclusions and avoid the limitations of "black box" algorithms. By using XAI, clinicians can make more informed decisions with greater confidence.

\subsection{Edge, fog, and cloud computing for implemetation} 
The edge network is a combination of edge computing and AI that processes algorithms based on AI near the source of data \cite{sayed2023edge}. This allows for better performance and lower costs for applications that require heavy information processing, as well as reduces the need for long-distance communication between the patient and the doctor. The proximity of the information and storage capabilities to the end-user in the health sector allows for direct and immediate access \cite{alsalemi2022innovative}. To further enhance performance, the detection of thyroid cancer in edge networks relies on the use of fog computing, which is a decentralized computing architecture located between the cloud and the data-producing devices. This architecture allows for the flexible placement of computing and storage resources in logical locations, improving performance \cite{sayed2021intelligent}.
To ensure the proper operation of the AI-based thyroid cancer detection system, it utilizes cloud computing as an access point. This guarantees that the stored data, servers, databases, networks, and programs are accessible and shared among specialized doctors, as long as it is connected to the Internet.
Such a hybrid system has proven to be effective for medical applications, including detection of thyroid cancer, as seen in various studies including \cite{charteros2020edge, sufian2020survey, chen2018disease, chai2020diagnosis, jagtap2016online, anuradha2021iot, kevco2018cloud, rajan2020fog, mutlag2019enabling, hartmann2019edge, corchado2019ai}.

\subsection{Reinforcement learning (RL)}
\begin{figure*}[t!]
\centering
\includegraphics[width=1.8\columnwidth]{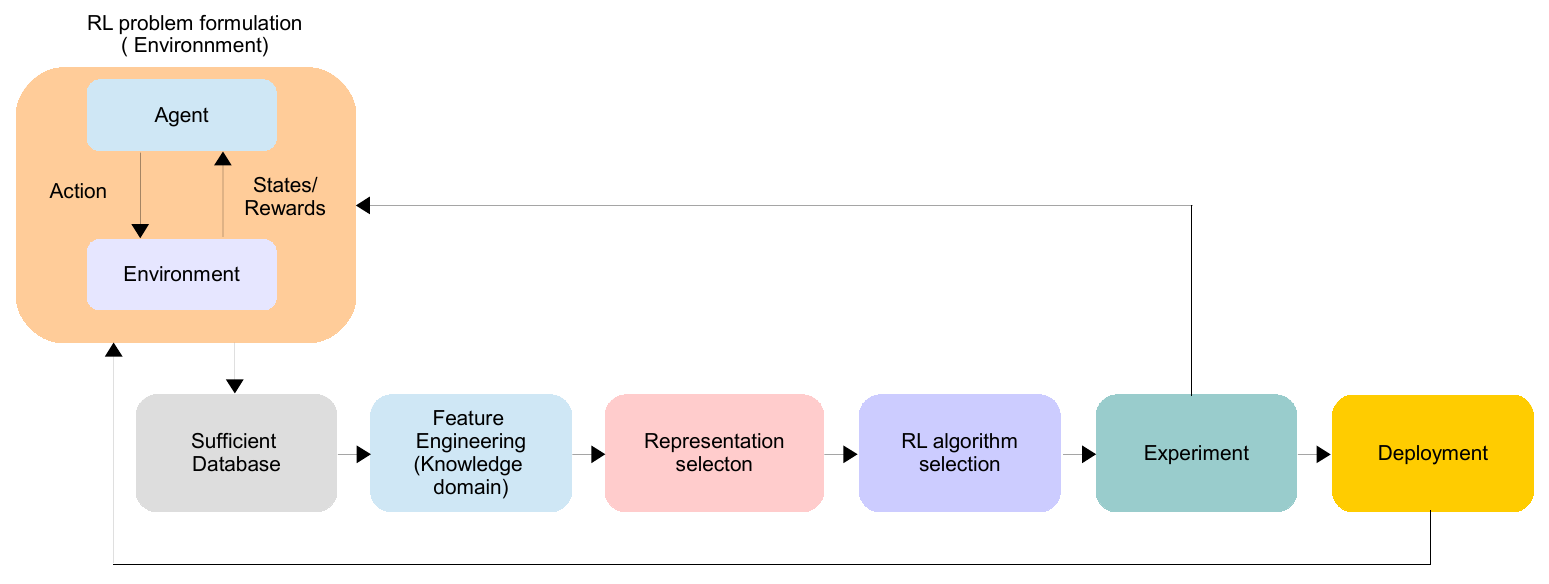}
\caption{Deep reinforcement learning procedure.}
\label{fig9}
\end{figure*}

RL, a subfield of ML, allows agents to make decisions in interactive environments through trial and error, observation, and learning (as depicted in Fig. \ref{fig9}). In recent years, there has been significant interest in using RL in detecting incurable diseases and providing explanations to aid medical decision-making. For example, reinforcement learning is used in \cite{balaprakash2019scalable} to classify cancer data, and deep reinforcement learning is used in \cite{li2020deep} to segment lymph node sets. The authors generate pseudo-ground truths using RECIST-slices and achieve simultaneous optimization of lymph node bounding boxes through the interaction between a segmentation network and a policy network.

\subsection{Transfer learning (TL)}
\begin{figure*}[t!]
\centering
\includegraphics[width=1.8\columnwidth]{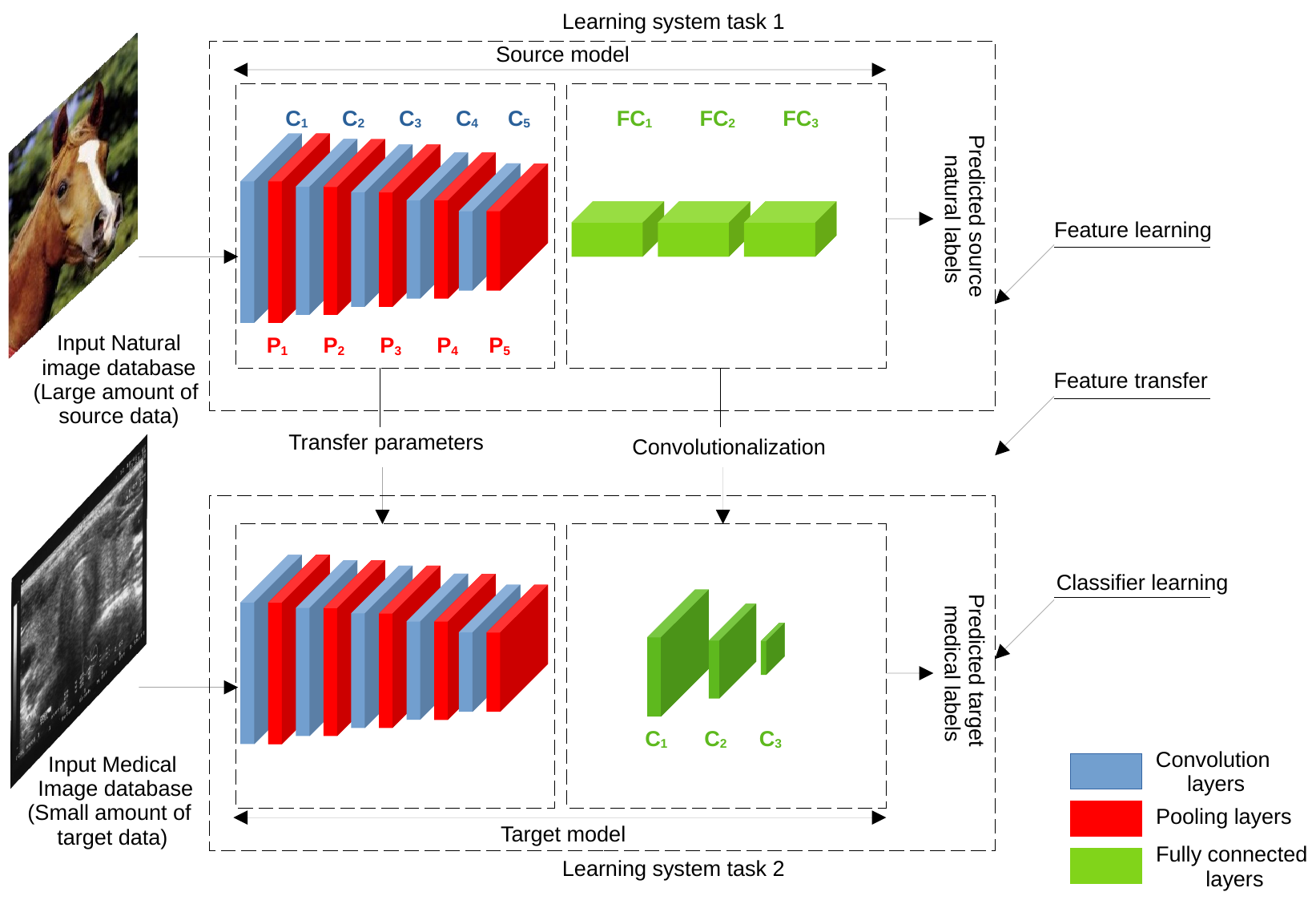}
\caption{Deep transfer learning.}
\label{fig10}
\end{figure*}

TL is a valuable solution to the overfitting and precision challenges faced by diagnosis systems \cite{kerdjidj2023uncovering,kheddar2023deep,himeur2023video}. This technique leverages stored knowledge from a specific problem to address other issues such as reducing training time and data volume \cite{kheddar2023deep,sayed2023time}. Its use in the diagnosis of the thyroid gland (TG) is demonstrated in Fig. \ref{fig10}. Moving on, in \cite{liu2017classification}, the authors tackle the challenge of capturing appropriate features of benign and malignant nodules using CNNs. They transfer the knowledge learned from natural data to an ultrasound (US) image dataset to produce hybrid semantic deep features. The transfer learning technique has also been successfully applied to classify thyroid nodule (TN) images in \cite{song2019ultrasound}. Other related works can be found in \cite{yu2020lymph,kwon2020ultrasonographic, wang2019using,ma2019thyroid,sundar2018exploring}.

\subsection{Panoptic segmentation (PS)}
\begin{figure*}[t!]
\centering
\includegraphics[width=1.6\columnwidth]{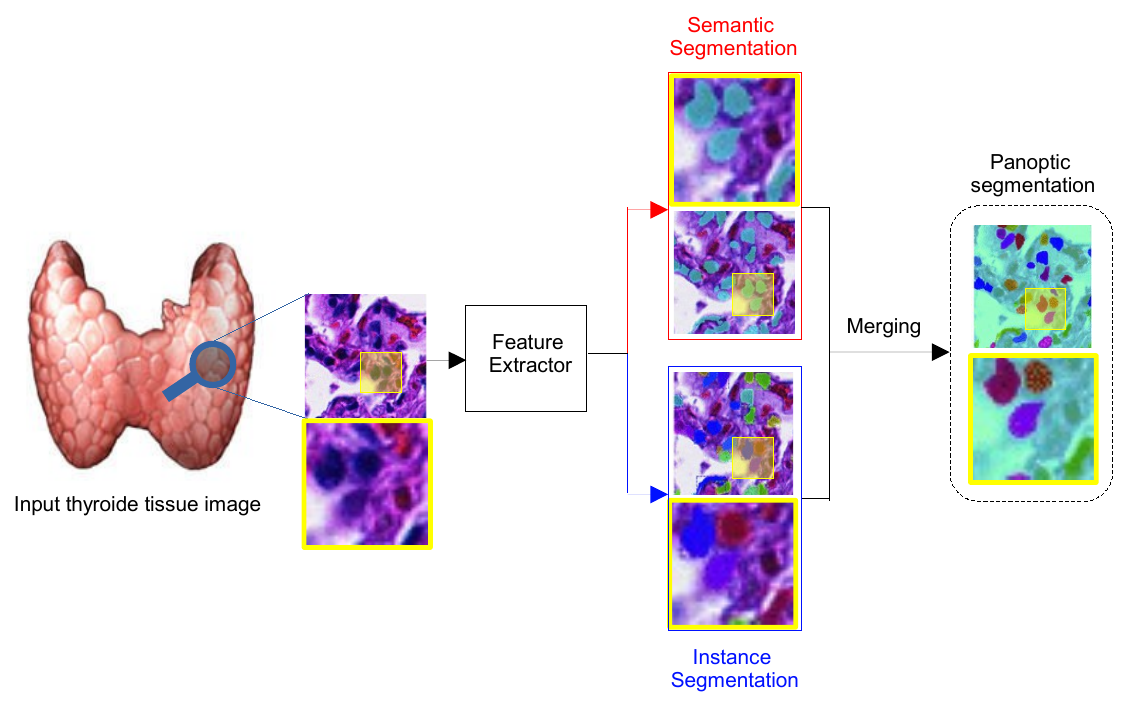}
\caption{Panoptic segmentation architecture.}
\label{fig11}
\end{figure*}

The challenge of accurately separating and dividing objects with diverse and overlapping appearances remains an issue, particularly in the medical field. To address this, many researchers have put forth proposals for a comprehensive and cohesive segmentation of various details \cite{liu2019nuclei,elharrouss2021panoptic}. The focus has been on PS, which combines both instance and semantic segmentation to identify and separate objects. In semantic segmentation, the goal is to classify each pixel into specific classes, while in instance segmentation, the focus is on segmenting individual object instances. AI has been incorporated into this model through supervised or unsupervised instance segmentation learning, making it well-suited for medical applications (Fig. \ref{fig11}). This has been demonstrated in works such as \cite{yu2020deep,cai2020panoptic}.

\subsection{Internet of medical imaging thing (IoMIT)} 
\begin{figure*}[t!]
\centering
\includegraphics[width=1.6\columnwidth]{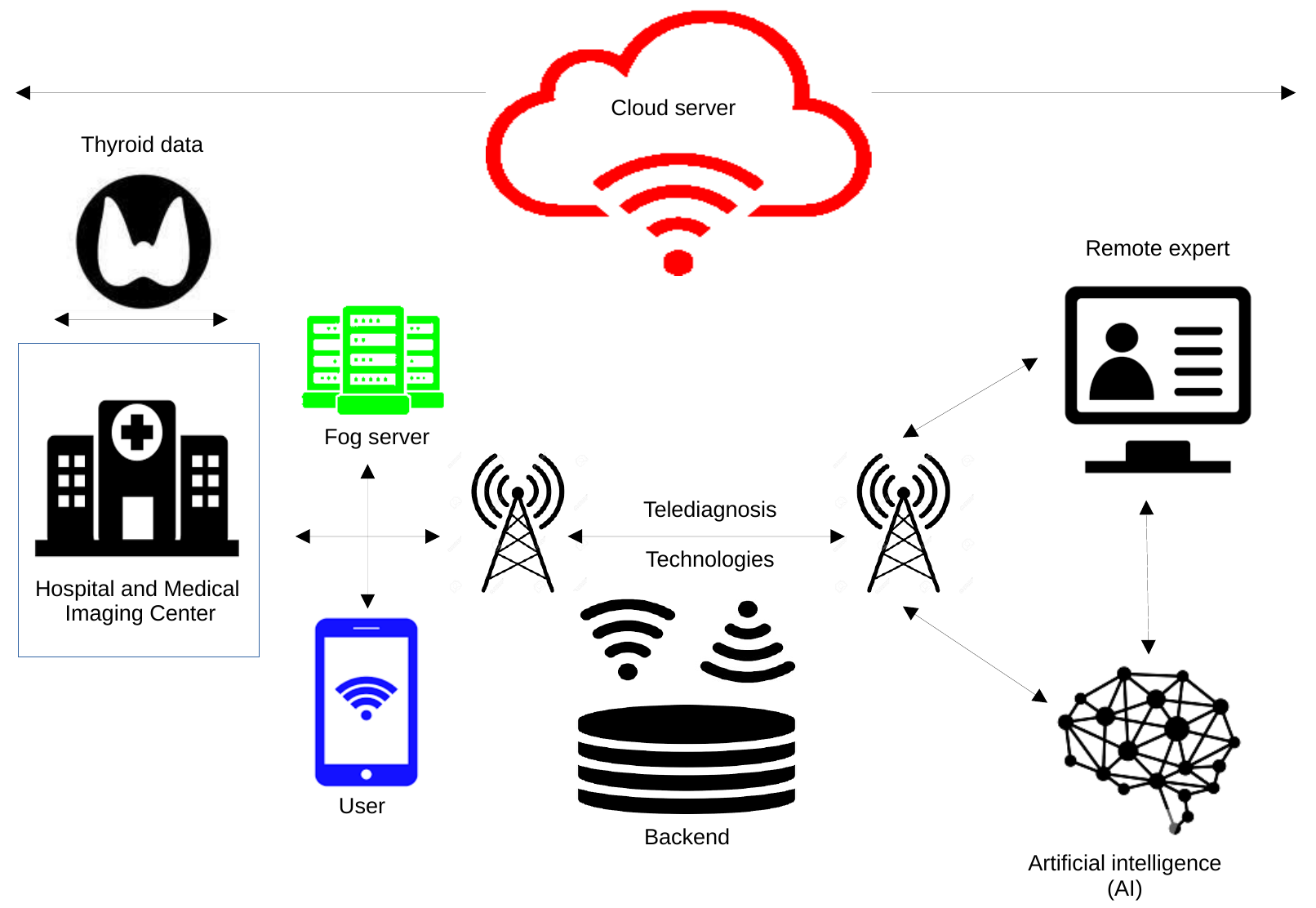}
\caption{Example of hybrid networks system based on AI for thyroid cancer detection.}
\label{fig12}
\end{figure*}

The Internet of Medical Things (IoMT) has recently gained widespread attention in the medical field, as it seeks to enhance healthcare delivery and reduce treatment costs through the exchange of health data between patients and doctors using connected devices with wireless communication (Fig. \ref{fig12}). One example of this integration can be found in \cite{ivanova2018artificial}, which proposes an AI-based solution for early detection of thyroid cancer in the IoMT, utilizing CNN to improve differentiation between benign and malignant nodules, ultimately saving lives. Other relevant studies related to the IoMIT have also been conducted, such as \cite{borovska2018internet} and \cite{ivanova2017internet}.

\subsection{3D thyroid cancer detection (3D-TCD)}
The conventional 2D ultrasound is widely used for diagnosing thyroid nodules, but its static images may not accurately reflect the nodule's structures. Hence, the use of 3D ultrasound has gained attention as it provides a more comprehensive view of the lesion by reconstructing its features and enabling better differentiation between different diagnoses \cite{seifert2023optimization}. With the ability to examine complex growth patterns, and margins, and to shape from multiple angles and levels, 3D ultrasound can provide a more accurate evaluation of the morphological features of thyroid nodules in comparison to 2D images. This has been confirmed through comparative studies between 3D and 2D ultrasound images \cite{li2015comparison, lyshchik2004accuracy, ying2008two}.

\subsection{AI in Thyroid Surgery (AI-TS)}
In light of the challenges faced in surgical procedures, the use of AI-powered robots in surgical practices is becoming increasingly essential. AI has the potential to address numerous clinical issues by analyzing and sharing massive amounts of data to support decisions with a level of accuracy comparable to that of healthcare professionals \cite{pakkasjarvi2023artificial}. Companies are incorporating AI into surgical practices by training AI-based systems and providing robots that assist surgeons in operating rooms, supply surgical materials, handle contaminated materials and medical waste, remotely monitor patients, and collect and organize patient data such as electronic medical records, vital signs, laboratory results, and video footage \cite{bodenstedt2020artificial}. As such, it is important for surgeons to have a strong understanding of AI in order to grasp its impact on healthcare. While AI-powered robotic surgery may still be some time away, collaboration across various fields can accelerate AI's capabilities and improve surgical care \cite{zhou2019artificial,zhou2020application,berikol2020artificial,tan2020part} \cite{habal2020brave,lee2020evaluation,voglis2020feasibility,navarrete2020current}.

\subsection{Wavelet-based AI} 
\begin{figure*}[t!]
\centering
\includegraphics[width=1.6\columnwidth]{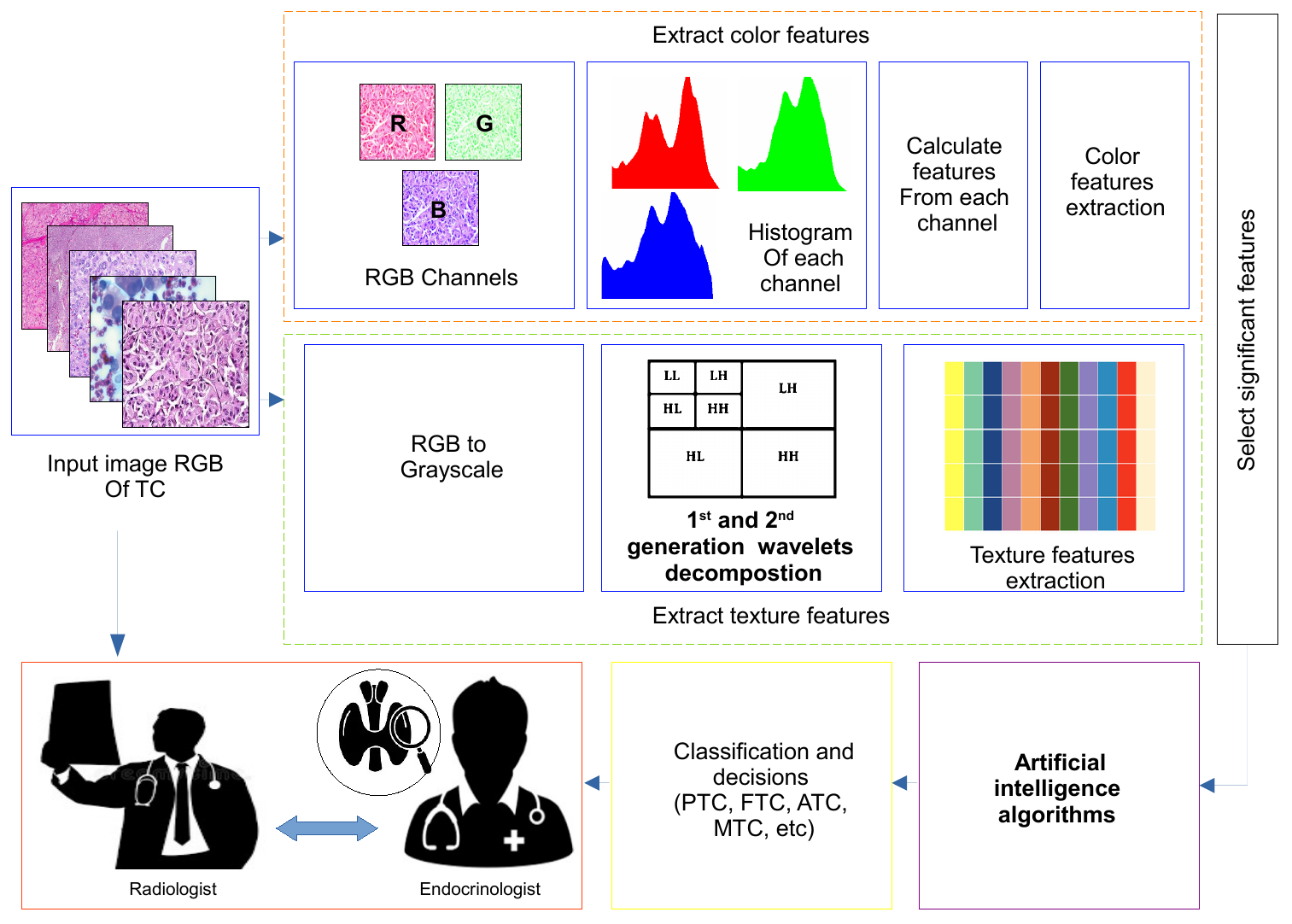}
\caption{Applications of AI-based on wavelet in the detection of thyroid cancer.}
\label{fig13}
\end{figure*}

Recently, wavelets transform, specifically first and second-generation, has gained recognition for its ability to detect various forms of cancer, especially when integrated with AI. This combination has become crucial in the medical field, providing doctors and surgeons with a tool to accurately diagnose diseases more efficiently and quickly \cite{melarkode2023ai,kumar2022artificial}. The proposed method is based on pre-processing the dataset through discrete wavelet transfer (DWT) and then evaluating the performance of AI in classifying different types of tumors that can impact organs in the body (as explained in Fig. \ref{fig13}). This model holds great potential for the detection of thyroid cancer and researchers are encouraged to test different wavelets available in the literature to further improve its effectiveness \cite{abdolali2020systematic}.

\subsection{Learning with reduced data} 
\begin{figure}[t!]
\centering
\includegraphics[width=1\columnwidth]{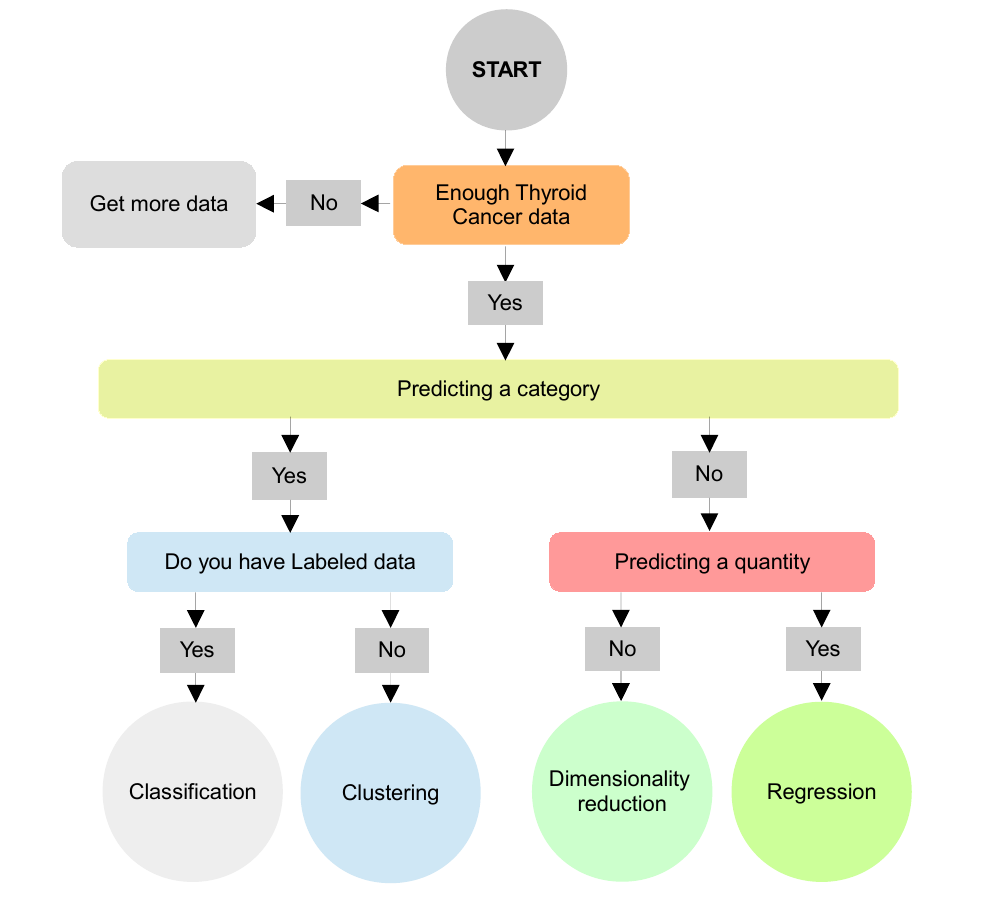}
\caption{Diagram of AI-algorithms choice for thyroid cancer detection.}
\label{fig14}
\end{figure}

One of the hurdles in implementing AI in the medical sector is acquiring adequate data and annotations. AI's capability to minimize the need for labeled data in making an accurate diagnosis is crucial \cite{petersson2022challenges}. This can be achieved through various learning methods such as semi-supervised learning, supervised learning, unsupervised learning, or alternative approaches that necessitate a smaller quantity of annotated data (Fig. \ref{fig14})\cite{sarker2022ai}.

\subsection{Recommender systems (RSs)}
\begin{figure*}[t!]
\centering
\includegraphics[width=1.6\columnwidth]{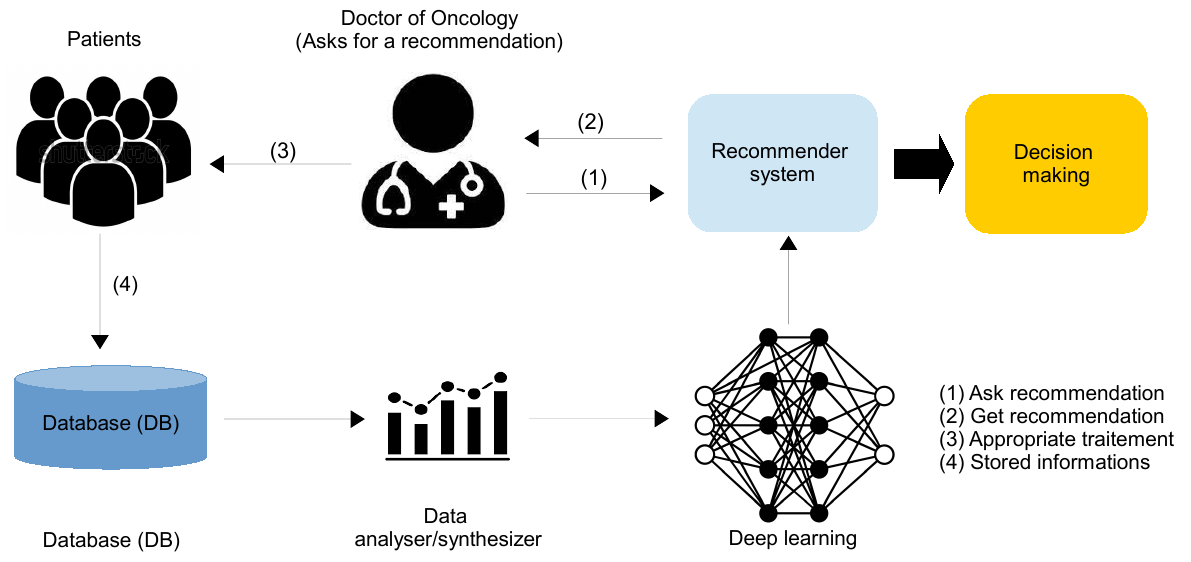}
\caption{Recommender systems for thyroid cancer detection.}
\label{fig15}
\end{figure*}

The abundance of data collected from online medical platforms and electronic health records can make it challenging for thyroid cancer patients to access relevant and accurate information \cite{himeur2021survey}. The high cost of healthcare data also poses difficulties for doctors to track patients and manage a large patient volume with various treatment options. Given these challenges, the implementation of RSs has been proposed to improve decision-making in healthcare and ease the workload for both patients and oncologists \cite{areeb2023filter,varlamis2022smart}. The use of RS in digital health provides personalized recommendations, accurate analysis of big data, and stronger privacy protection through integration with artificial intelligence and machine learning technologies \cite{atalla2023intelligent} as depicted in Fig. \ref{fig15}.

\subsection{Federated learning (FL):} 
\begin{figure}[t!]
\begin{center}
\includegraphics[width=0.95\columnwidth]{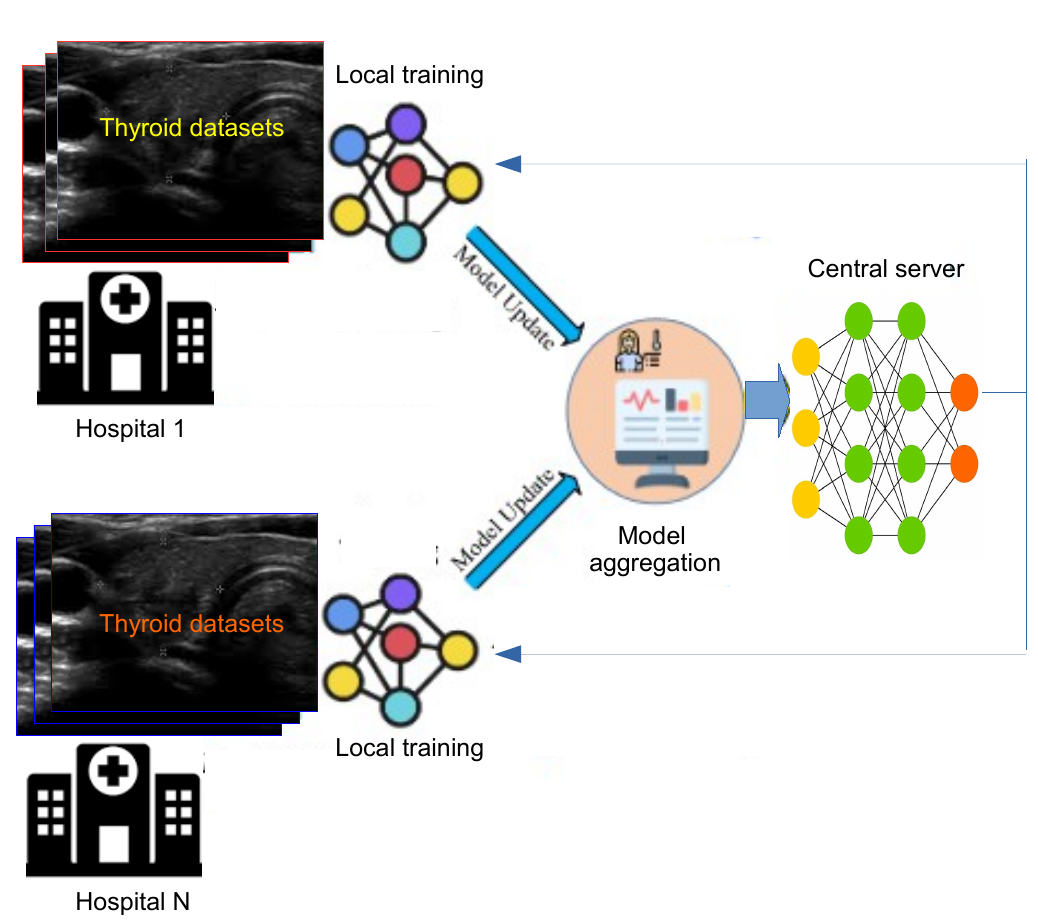}
\end{center}
\caption{The FL for healthcare .}
\label{fig17}
\end{figure}

The FL has become very popular in the field of healthcare applications \cite{antunes2022federated}. The surrounding conditions greatly affect human health and cause negative effects on the economy. Diseases of the thyroid gland are among the most common health problems that have become noticeable among various groups of society in recent times.
ML can play a vital role in such medical conditions, as the collected data can be exploited to train an ML model that can predict critical conditions. Emphasizing that patient data across different medical centers should be handled privately, the FL setup is the natural choice for such applications, as depicted in Fig.\ref{fig17}. Therefore, In \cite{lee2021federated}, the authors compared the performance of FL against the five conventional deep learning (VGG19, ResNet50, ResNext50, SE-ResNet50, and SE-ResNext50) for analysing and detect thyroid cancer datasets.

\subsection{Generative chatbots}
Most recently, the realm of artificial intelligence has witnessed significant advancements, particularly in the development of generative chatbots and large language models like GPT variants \cite{sohail2023using}. These state-of-the-art models, trained on vast amounts of data, are adept at generating human-like text and engaging in coherent conversations, going beyond mere predefined responses. As their capability has expanded, so too has their potential for application across various domains, healthcare being one of the prominent ones. In the healthcare sector, these sophisticated models are being explored for patient engagement, preliminary symptom checks, providing health-related information, and even assisting professionals with medical research and data analysis \cite{sohail2023decoding}. The integration of such technology holds the promise of streamlining healthcare processes, enhancing patient experience, and augmenting the capabilities of healthcare professionals, albeit with the necessary precautions and ethical considerations in place \cite{farhat2023analyzing}.

Using generative chatbots or models like ChatGPT to diagnose thyroid cancer (or any medical condition) directly would be inappropriate and potentially dangerous. However, they can be incorporated into healthcare settings in auxiliary roles \cite{sohail2023future}. Typically, Chatbots can gather preliminary information from patients, including their symptoms, family history, and lifestyle habits. This data can provide a better understanding of the patient's concerns before they meet a healthcare professional. Moving on, they can be programmed to provide information about thyroid cancer, such as risk factors, symptoms, and preventive measures \cite{haver2023appropriateness}. Patients can learn about the disease and its potential signs, allowing them to approach healthcare providers if they find any matching symptoms. Besides, while they cannot replace professional diagnostic tools, they can be designed to guide users through a series of questions that could highlight potential risk factors or symptoms, encouraging them to consult a medical professional for a more comprehensive evaluation \cite{cao2023accuracy}.

On the other hand, once a diagnosis has been made, chatbots can provide patients with information on treatment options, side effects, diet recommendations, and answer frequently asked questions. Additionally, they can (i) remind patients to take their medications, attend follow-up appointments, or perform regular self-examinations or monitoring, (ii) offer support in terms of relaxation techniques, provide resources for further psychological support, or even just offer a non-judgmental "listening ear", and (iii) assist doctors and other healthcare professionals by providing instant information about thyroid cancer, recent research, or treatment options, acting as a dynamic reference tool \cite{sorin2023large}.

\section{Conclusion}
In this research, a comprehensive overview of Deep Neural Networks (DNNs) has been presented, highlighting their growing trend in recent years due to their high accuracy in results compared to other methods. A range of algorithms and training structures have been described, including their advantages and limitations. DNNs have been shown to play a critical role in various real-world applications, characterized by their generalizability and noise tolerance.

However, there are still challenges to be addressed for the widespread use of DNNs in the detection of Thyroid Cancer. One such challenge is the lack of clean datasets and platforms. It is crucial to consider these data to develop efficient and robust cancer detection models that can identify more advanced types of cancer. In the future, more research effort should be put into overcoming these problems and improving the quality of thyroid cancer detection.

The study also highlights the need for further research in the area of thyroid cancer detection and identification, especially considering the accuracy that health specialists desire. Researchers are increasingly exploring the detection of various cancers in two or three dimensions, but the lack of mastery of different geometric transformations and a two- or three-dimensional database hinders the process of diagnosing incurable diseases. The development of new methods to recognize different volumes of cancerous nodules is crucial to achieving speed in treatment and accuracy in diagnosis, as well as enabling early epidemiological surveillance and reducing the death rate.

New technologies such as Explainable AI, Edge Computing, Reinforcement learning, Panoptic Segmentation, and Recommender Systems, have opened up new avenues for research in the field of thyroid cancer detection and have greatly assisted clinicians in the early diagnosis process, reducing the time for detection, and preserving patient privacy. Future work will focus on further exploring the contributions of these technologies to drive a paradigm shift in the field of cancer detection, by developing advanced and secure technologies for the preservation of privacy and detection of thyroid cancer patients, such as telehealth.



\end{document}